\documentclass[aps,prb,reprint,amsmath,amssymb,groupedaddress]{revtex4-2}

\usepackage{graphicx}
\usepackage{dcolumn}
\usepackage{bm}
\usepackage{color}
\usepackage{siunitx}
\usepackage{caption}
\usepackage{subcaption}
\usepackage{todonotes}
\usepackage{threeparttable}
\usepackage{gensymb}
\usepackage{natbib}
\usepackage{url}
\usepackage{textcase}

\begin{document}


\title{Oxygen vacancy induced site-selective Mott transition in LaNiO$_{3}$}


\author{Xingyu Liao$^{1}$, Vijay Singh$^{1}$, and Hyowon Park$^{1,2}$}
\affiliation{$^1$Department of Physics, University of Illinois at Chicago, Chicago, IL 60607, USA \\
$^2$Materials Science Division, Argonne National Laboratory, Argonne, IL, 60439, USA }


\date{\today}

\begin{abstract}
While defects such as oxygen vacancies in correlated materials can modify their electronic properties dramatically, understanding the microscopic origin of electronic correlations in materials with defects has been elusive.
Lanthanum nickelate with oxygen vacancies, LaNiO$_{3-x}$, exhibits the metal-to-insulator transition as the oxygen vacancy level $x$ increases from the stoichiometric LaNiO$_3$.
In particular, LaNiO$_{2.5}$ exhibits a paramagnetic insulating phase, also stabilizing an antiferromagnetic state below $T_N\simeq152$K.
Here, we study the electronic structure and energetics of LaNiO$_{3-x}$ using first-principles.
We find that LaNiO$_{2.5}$ exhibits a ``site-selective" paramagnetic Mott insulating state at $T\simeq290$K as obtained using density functional theory plus dynamical mean field theory (DFT+DMFT).
The Ni octahedron site develops a Mott insulating state with strong correlations as the Ni $e_g$ orbital is half-filled while the Ni square-planar site with apical oxygen vacancies becomes a band insulator.
Our oxygen vacancy results cannot be explained by the pure change of the Ni oxidation state alone within the rigid band shift approximation.
Our DFT+DMFT density of states explains that the peak splitting of unoccupied states in LaNiO$_{3-x}$ measured by the experimental X-ray absorption spectra originates from two nonequivalent Ni ions in the vacancy-ordered structure.

\end{abstract}

\maketitle

\section{Introduction}

Rare-earth nickelates $R$NiO$_3$ ($R$ is the rare-earth ion) have attracted significant research interests due to their rich electronic properties. These include the metal-insulator transition, charge order, magnetism, multiferroicity, and the site-selective Mott transition~\cite{rev_RNO,rev_RNO_2,RNO_charge_order,NNO_charge_order,RNO_mag_trans,PhysRevLett.109.156402}. 
Although the $R$ ion can be treated as electrically inert, the phase boundaries of the metal to insulator transition and the paramagnetic (PM) to anti-ferromagnetic (AFM) transition depend sensitively on the subtle structural tolerance factor controlled by the size of the $R$ ion~\cite{RNO_MIT_rev}.
This close interplay between the structural, electronic, and magnetic degrees of freedom puts the rare-earth nickelate into an intriguing correlated material.

Oxygen vacancy is one of the common defects in transition metal oxides and it can play a central role in oxide electronics~\cite{doi:10.1063/1.5143309}. 
It is also known that oxygen vacancies in LaNiO$_3$, one of the rare-earth nickelates, also change
its electronic and magnetic properties significantly as they can modify electronic correlation effects.
Although LaNiO$_3$ is the only metallic case among the known rare-earth nickelate series, 
Ni $d$ orbitals are still moderately correlated as indicated by experimental spectroscopic measurements~\cite{PRB.79.115122,PRB.83.075125,PRB.82.165112}.
Experimental measurements on the conductivity in LaNiO$_{3-x}$ show that the increase of the vacancy level $x$ reduces the conductivity and the metal-to-insulator transition occurs as $x$ approaches to 0.5~\cite{MORIGA1995252_LNO2.6,Sanches_MIT}.
As the oxygen vacancy level $x$ increases further, LaNiO$_{2+\delta}$ is found to be semiconducting or poorly conducting~\cite{LNO2_Metallic}. 
The complete absence of the apical oxygens leads to the infinite-layer structure of LaNiO$_2$ and the role of electronic correlations in LaNiO$_{2}$ has been drawing much attention recently as similar nickelates such as NdNiO$_{2}$ and PrNiO$_{2}$ exhibit superconductivity when they are hole-doped~\cite{nature_NdNiO2_sc,PrNiO2}.
Although LaNiO$_{2}$ is metallic, resistivity increases at low temperatures hinting possibly strong correlation effects.

In addition to transport properties, oxygen vacancies also have significant effects on magnetism. 
Although LaNiO$_3$ has been known to remain PM at all temperatures, there was a controversial experimental work showing some evidence of AFM orders in LaNiO$_3$~\cite{LNO3_afm}. It was also argued that AFM in LaNiO$_3$ can be originated from small oxygen vacancies~\cite{LNO3_PM2,LNO_FM_AFM_2018}. 
As oxygen vacancy level further increases, LaNiO$_{2.5}$ becomes AFM below 152K, and LaNiO$_{2.75}$ shows ferromagnetic (FM) structure below 225K~\cite{LNO_FM_AFM_2018}. 
However, LaNiO$_{2}$ does not show any clear evidence of the long-range magnetic order~\cite{LNO2_PM}.

Spectroscopic measurements of LaNiO$_{3-x}$ are also widely performed using X-ray Absorption Spectroscopy (XAS) and  Photo-Emission Spectroscopy (PES) to study electronic structure in experiments.
Consistently with the transport measurement, spectra at the Fermi energy decreases as the vacancy level $x$ increases from LaNiO$_3$, opening a spectral gap near the level at $x=0.5$.
An interesting feature measured from XAS in LaNiO$_{3-x}$ bulk~\cite{XAS,LaNiO3_PES_XAS} as well as the thin-layer structure~\cite{Natcomm18} is the splitting of the spectral peak above the Fermi energy, which has been attributed to the oxygen vacancy effect. 

There have been some first-principles studies of magnetism and oxygen vacancy effects on rare-earth nickelates.
Previous density functional theory (DFT) and $GW$ study on LaNiO$_{3-x}$ systems addressed the metal-insulator transition and resulting spectra due to the vacancy effect~\cite{EPJB16}.
A. Malashevich $et$ $al$~\cite{DFT_study} did a systematic study on LaNiO$_{3-x}$ with small $x$ value and found that oxygen vacancies stay around the same Ni ion and localize extra electrons created by the vacancy. 
The strong localization of electrons due to the oxygen vacancies in other rare-earth nickelates also has been studied using DFT+U~\cite{Kotiuga21992}.
A. Subedi $et$ $al$~\cite{10.21468/SciPostPhys.5.3.020} studied structural and magnetic instabilities in LaNiO$_{3}$ with possible breathing-type lattice distortions using DFT and DFT+U.
Theoretical studies on $R$NiO$_{2}$ with $R$=La, Nd, Sr, and Pr have attracted much attention recently as they can serve as model systems of experimentally discovered nickelate superconductors~\cite{NNO_PNO_LNO_sc,LNO_CCO_sc,PhysRevX.10.021061, ryee2019induced, werner2019nickelate, lechermann2019late, PhysRevB.102.161118}.
Nevertheless, the microscopic origin of the strongly correlated insulating phase induced by oxygen vacancies and the changes of the correlated spectra in LaNiO$_{3-x}$ compared to experiments have not been systematically investigated. 

In this paper, we study the strong correlation effect on the electronic structure and the energetics of LaNiO$_{3-x}$ from first-principles as the oxygen vacancy level $x$ evolves.
We adopt dynamical mean field theory (DMFT) in combination with DFT to treat strong correlations in the paramagnetic phase as well as DFT+U for the long-range magnetic state.
We show that the vacancy-ordered structure becomes thermodynamically stable in LaNiO$_{2.5}$ and the metal-to-insulator transition due to the change of the vacancy level $x$ can be captured correctly in DFT+DMFT. 
The insulating nature of the vacancy-ordered LaNiO$_{2.5}$ structure with two nonequivalent Ni ions originates from the site-selective Mott phase due to both structural and electronic correlation effects.
While bulk LaNiO$_3$ forms a rhombohedral structure with the octahedral geometry of the Ni ion surrounded by six O ions, oxygen vacancies can change both the oxidation number of the Ni ion and the local structure, which can lead to the substantial change of electronic structures in LaNiO$_{3-x}$.

Our paper is organized as follows.
First, we explain the computational methods we used including DFT, DFT+U and DFT+DMFT in Sec.$\:$\ref{sec:method} and show the structural details and magnetism in Sec.$\:$\ref{sec:structurals}. 
We also study formation energies of LaNiO$_{3-x}$ in Sec.$\:$\ref{sec:energetics}. 
Then we display the spectral functions of LaNiO$_{3-x}$ computed using DFT+DMFT and DFT+U, and compare to experimental measurements in Sec.$\:$\ref{sec:spectra}.
The DMFT self-energies in LaNiO$_{3-x}$ are displayed to explain the nature of the insulating phase in Sec.$\:$\ref{sec:selfenergy} and compare our results to the rigid band shift approximation in Sec.$\:$\ref{sec:rigidband}.
And we conclude our discussion in Sec.$\:$\ref{sec:conclusion}.

\section{\label{sec:method}Computational Methods}

First, we performed structural relaxation calculations for LaNiO$_{3-x}$ ($x$=0, 0.25, 0.5, 0.75, and 1) systems and obtained the ground-state energies and magnetism using both DFT and DFT+U. Vienna Ab-initio Simulation Package (VASP)~\cite{vasp1,vasp2} has been used in all DFT and DFT+U calculations adopting the Perdew-Burke-Ernzerhof for solids (PBE-sol)~\cite{PBEsol} as the exchange and correlation energy functional. We set 600 eV as the energy cutoff for the plane wave basis and a Gaussian smearing of 0.2 eV is used for the summation over the Monkhorst-Pack $k-$point mesh. For LaNiO$_3$, LaNiO$_{2.5}$ and LaNiO$_2$ structures, we use a 8$\times$8$\times$8 $k-$point grid. LaNiO$_{2.25}$ and LaNiO$_{2.75}$ have a rather large supercell with a long lattice vector along the $y-$direction, therefore we use a 6$\times$3$\times$6 $k-$grid. For all structural relaxations, we set 0.001 eV/\AA $\:$ as the force convergence condition fully relaxing the cell shape, volume and internal ionic positions. 

Then we calculate the correlated electronic structure using DFT+DMFT and DFT+U for LaNiO$_3$, LaNiO$_{2.5}$ and LaNiO$_2$. 
While DFT+U can capture static correlations beyond DFT at the Hartree-Fock level based on single-determinant wavefunctions, DFT+DMFT can go beyond the static approximation in DFT+U and capture dynamical correlations based on multi-determinant many-body wavefunctions.    
To study magnetism, DFT+U is adopted to relax structures imposing experimental magnetic orderings and to study the electronic structure from those relaxed structures. 
For a paramagnetic state, we adopt DFT+DMFT using the DFT relaxed structure with the non-magnetic(NM) order. 
The relaxed structures obtained using DFT and DFT+U are quite similar, as will be shown in next section.
We adopt the DMFTwDFT package~\cite{SINGH2020107778} for DFT+DMFT calculations. 
Wannier90 package~\cite{MLWF1,MLWF2} has been adopted to obtain maximally localized Wannier functions for the construction of the DMFT correlated subspace.
Nickelates show a rather strong $d-p$ hybridization due to the covalent bonding between Ni and O ions.
Therefore, it is important to construct both Ni 3$d$ and O 2$p$ orbitals for the Wannier basis to treat the hybridization effect.
To construct the Wannier orbitals, we take an energy window from -9 eV to 5 eV from the Fermi energy, which basically all Ni 3$d$ and O 2$p$ orbitals.  

For DFT+U and DFT+DMFT, we need to define interaction parameters to treat the on-site Coulomb interaction within the $d-$orbitals of the Ni ion. A rather small value of $U$($\simeq$2eV) was used in the $d-$orbital model of previous rare-earth nickelates studies~\cite{PhysRevB.92.155145,Hampel19} while $U$=5$\sim$7eV was used for the wide-energy window calculation including both $d-$ and $p-$orbitals to reproduce the metal-insulator and structural phase diagram and to compare with the angle resolved photoemission spectra~\cite{phase_LNO3,ARPES_DMFT}. 
Similar $U$ value ($\simeq$5.7eV) was also obtained from the constrained DFT calculation using the Quantum ESPRESSO code~\cite{PhysRevB.84.144101}.
For both DFT+DMFT and DFT+U calculations in this paper, we use the Hubbard $U$=5eV and the Hund's coupling $J$=0.8eV which are parameterized by the Slater integrals.
To account for the double-counting correction of DFT+DMFT, the modified fully-localized-limit form of the double-counting potential, which was used for the phase diagram study of rare-earth nickelates~\cite{phase_LNO3,Park_dc_2014}, has been used.
To solve the DMFT impurity problem, we use the continuous-time quantum Monte Carlo~\cite{CTQMC,PhysRevB.75.155113} solver with temperature $T\approx$ 290K.
After DFT+DMFT calculations are converged, we used the post-processing tool from the DMFTwDFT package to calculate the spectral function A($\omega$) for LaNiO$_{3-x}$. 
More details of the DMFT calculation method are shown in the Supplemental Material.

\section{Results}
\subsection{\label{sec:structurals}Structural relaxation and magnetism}

\begin{figure}
    \begin{subfigure}[b]{0.47\linewidth}
       \includegraphics[width=\linewidth]{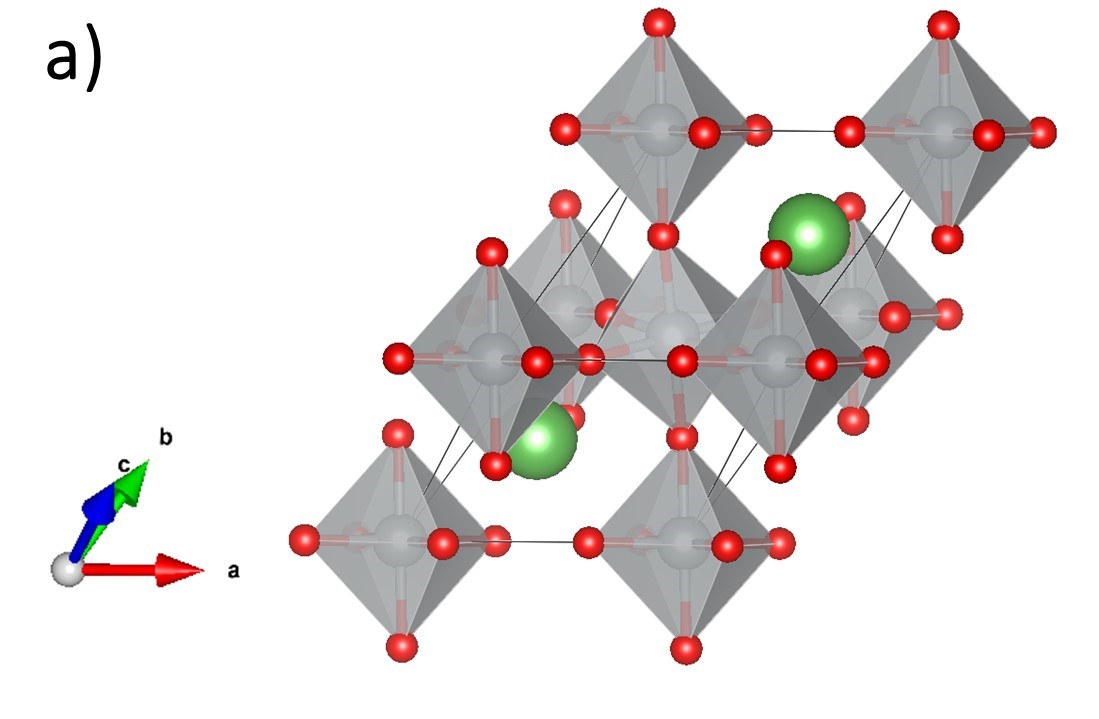}
    \end{subfigure}
    \begin{subfigure}[b]{0.47\linewidth}
        \centering
        \includegraphics[width=\linewidth]{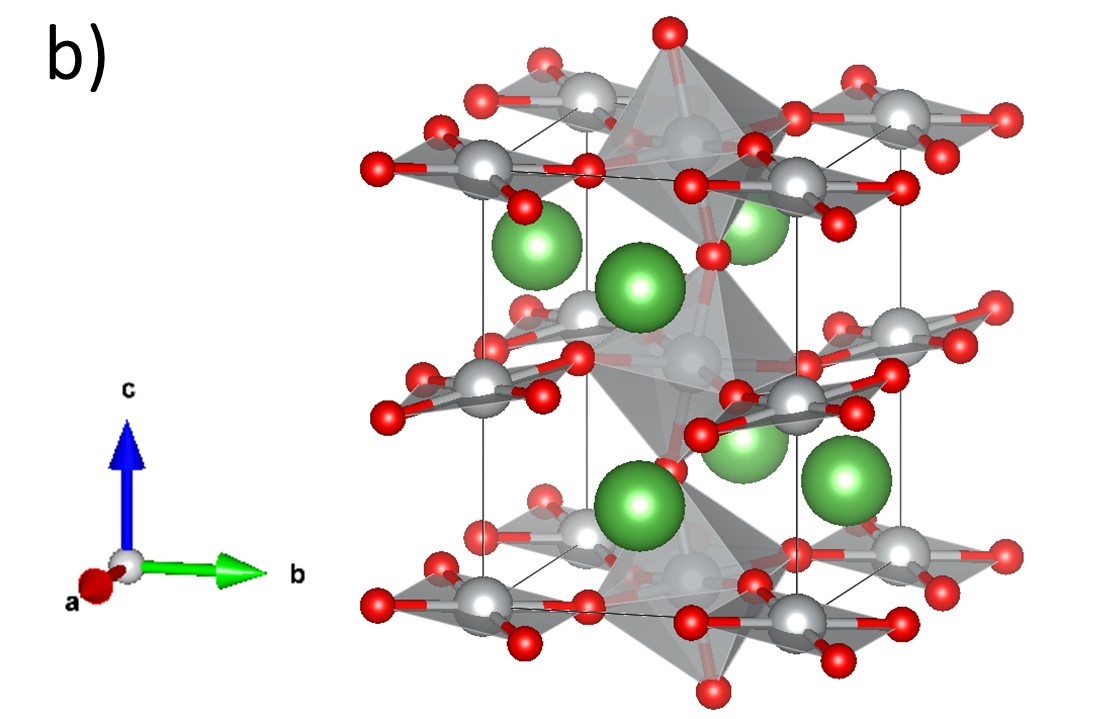}
    \end{subfigure}
    \begin{subfigure}[b]{0.47\linewidth}
        \centering
        \includegraphics[width=\linewidth]{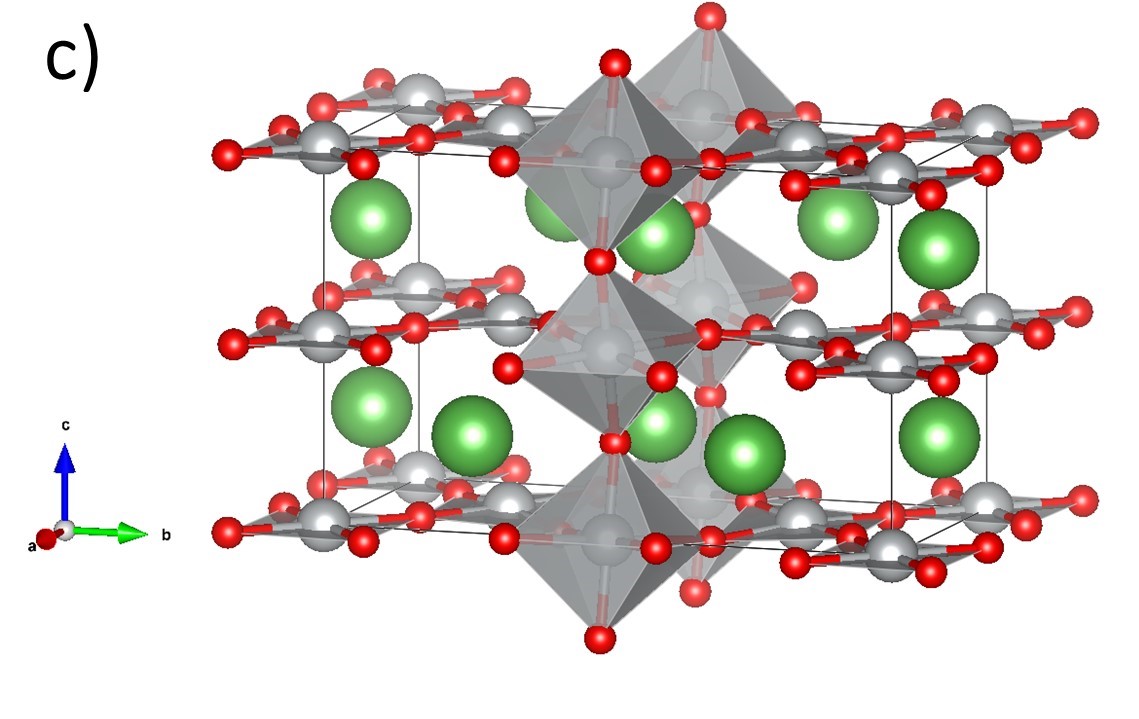}
    \end{subfigure}
    \begin{subfigure}[b]{0.47\linewidth}
       \centering
       \includegraphics[width=\linewidth]{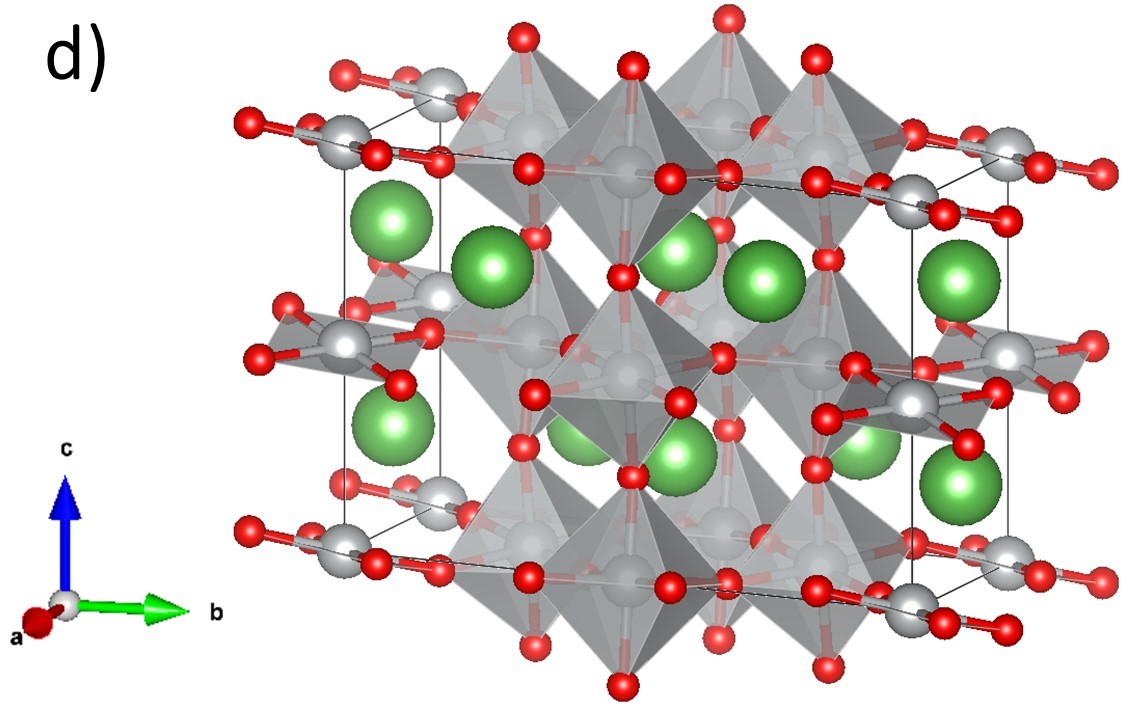}
    \end{subfigure}
    \begin{subfigure}[b]{0.47\linewidth}
        \centering
        \includegraphics[width=\linewidth]{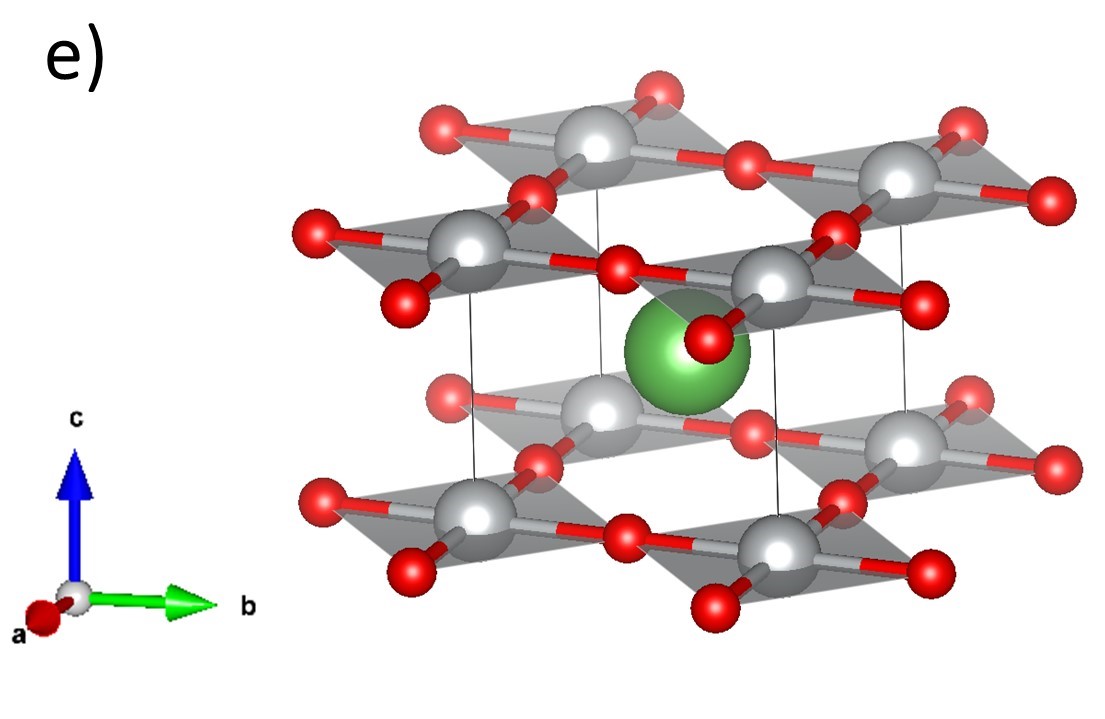}
    \end{subfigure}
    \begin{subfigure}[b]{0.47\linewidth}
        \centering
        \includegraphics[width=0.85\linewidth]{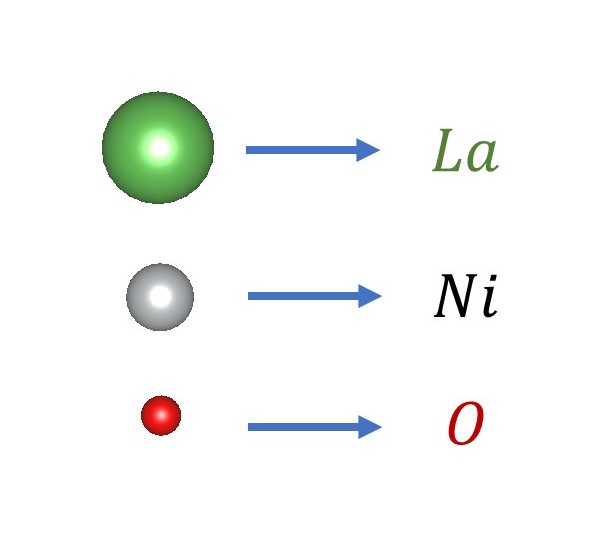}
    \end{subfigure}

\caption{\label{fig:structures}Crystal structures of (a) LaNiO$_3$, (b) LaNiO$_{2.5}$, (c) LaNiO$_{2.25}$, (d) LaNiO$_{2.75}$, and (e) LaNiO$_{2}$}
\end{figure}

Bulk LaNiO$_3$ forms a rhombohedral structure given by the $R\bar{3}c$ space symmetry group~\cite{LNO3_rhom}. This structure has two La, two Ni, and six O ions in a unit-cell, which can be obtained by rotating the Ni-O octahedra from the cubic perovskite structure. Namely, without any defects, each Ni ion is surrounded by six O ions forming the octahedron and all Ni-O octahedra are equivalent with the same Ni-O bond lengths (see Fig.$\:$\ref{fig:structures}a).
The oxygen vacancy formation induces the local structural distortion due to the absence of apical oxygens, which breaks the cubic symmetry. 
While it is challenging to measure the crystal structure with vacancies, several experiments~\cite{Alonso_1997,Sanches_MIT,Moriga_structure} suggest the LaNiO$_{2.5}$ structure such that NiO$_6$ octahedra and NiO$_4$ square-planes are alternating in the $x-y$ plane as shown in Fig$\:$\ref{fig:structures}b. 
Previous DFT calculation~\cite{DFT_study} also provides the insight that the apical divacancy configuration lowers the formation energy than other configurations meaning a four-coordinated Ni-O square plane is energetically favored when oxygen vacancies are introduced. 
In this paper, we denote Ni in the octahedral environment as Ni$_{o}$ and Ni in the square-planar symmetry as Ni$_{sp}$.
We also construct LaNiO$_{2.25}$ and LaNiO$_{2.75}$ structures with vacancy orderings such that Ni$_o$ and Ni$_{sp}$ ions modulated along the $x-y$ plane, as shown in Fig$\:$\ref{fig:structures}c and Fig$\:$\ref{fig:structures}d. 
LaNiO$_2$ becomes a tetragonal structure of purely Ni$_{sp}$ ions with NiO$_4$ square-planes (see Fig.$\:$\ref{fig:structures}e).

While LaNiO$_3$ remains a PM metallic state at all temperatures, LaNiO$_{3-x}$ ($x>0$) undergoes the magnetic transition for most cases.
In Table$\:$\ref{tbl:magnetic}, we list the experimentally observed long-range magnetic orderings of LaNiO$_{3-x}$ including the ground state (metal or insulator) and the Neel temperature ($T_N$) with relevant references. 
Most LaNiO$_3$ structures with O vacancies become FM except LaNiO$_{2.5}$.
Previous experimental work~\cite{Alonso_1997} in LaNiO$_{2.5}$ suggests that LaNiO$_{2.5}$ is the G-type AFM and the Ni$_o$ ions have relatively large magnetic moments, while the Ni$_{sp}$ ions have almost no magnetic moment. All LaNiO$_{3-x}$ structures become paramagnetic above $T_N$. 

To study the structural and magnetic properties, we fully relaxed the oxygen-vacancy ordered structures of LaNiO$_{3-x}$ with $x=$0, 0.25, 0.5, 0.75 and 1 using both DFT and DFT+U. 
In Table$\:$\ref{tbl:struc_thry_exp}, we provide the structure details obtained from the relaxations with magnetic structures observed in experiments for x=0, 0.5 and 1.
That is to say, we impose the G-type AFM order on LaNiO$_{2.5}$ relaxations. The non-magnetic(NM) order calculations on LaNiO$_{2}$ and LaNiO$_{3}$ are performed to simulate PM nature in experiment.
In LaNiO$_3$, DFT Ni-O bond length and Ni-O-Ni bond angle are similar to experiment while DFT+U overestimate the bond angle along with the contracted bond length.
LaNiO$_{2.5}$ has two nonequivalent Ni ions and the Ni$_o$-O bond length is much larger than the Ni$_{sp}$-O bond length. LaNiO$_{2.5}$ structure relaxed with DFT shows the Ni-O bond length difference to be 0.14\AA. DFT+U predicts the similar Ni-O bond difference ($\sim0.22$\AA) as the experimental value ($\sim0.21$\AA) although the absolute values of the bond lengths in DFT+U are smaller than experimental values.
In LaNiO$_{2}$, the Ni-Ni distance along $z-$axis is quite smaller than along the $x-y$ plane due to the loss of apical oxygens and the DFT structural parameters are closer to experimental values than the DFT+U parameters.

\begin{table}[ht]
    \centering
    \caption{\label{tbl:magnetic}
Experimental magnetic and transport (metal/insulator) properties of LaNiO$_{3-x}$} 
    \begin{tabular}{p{0.35\linewidth}p{0.18\linewidth}p{0.15\linewidth}p{0.1\linewidth}}
    \hline\hline
      LaNiO$_{3-x}$ & M/I & MAG. & T$_N$ \\ [1.0ex] 
    \hline
      LaNiO$_3$~\cite{LNO3_PM2,Sanches_MIT} & M & PM & 0K \\ 
      LaNiO$_{2.75}$~\cite{LNO_FM_AFM_2018,Sanches_MIT} & M & FM & 225K \\ 
      LaNiO$_{2.53}$~\cite{Moriga_structure} & N/A & FM & N/A \\ 
      LaNiO$_{2.5}$~\cite{LNO_FM_AFM_2018,Sanches_MIT} & I & AFM & 152K \\ 
      LaNiO$_{2}$~\cite{LNO2_Metallic,LNO2_PM} & M & PM & N/A \\ 
     \hline\hline
    \end{tabular}
\end{table}

\begin{table}[ht]
    \centering
    \caption{Structural information of LaNiO$_3$, LaNiO$_{2.5}$, and LaNiO$_2$ obtained from DFT and DFT+U relaxations.} 
    \begin{ruledtabular}
    \begin{tabular}{p{0.17\linewidth}|p{0.24\linewidth}p{0.12\linewidth}p{0.16\linewidth}p{0.13\linewidth}}
                  &Parameters       &DFT    & DFT+U & Exp\\
    \hline
      LaNiO$_{3}$ &d$_{Ni-O}$ [\AA]                 &1.90   &1.88   &1.93\cite{LNO_LattParaExp}  \\ 
       (NM)       &$\alpha_{Ni-O-Ni}$ [$^{\circ}$]  &164.5  &168.3  &164.8\cite{LNO_LattParaExp} \\
    \hline
    LaNiO$_{2.5}$ &d$_{Ni-Ni}$\footnotemark[1] [\AA]                  &3.83   &3.82   &3.91\cite{Alonso_1997}\\ 
        (AFM)     &d$_{Ni-Ni}$\footnotemark[2] [\AA]                  &3.64   &3.67   &3.74\cite{Alonso_1997}\\ 
                  &d$_{Ni_{sp}-O}$ [\AA]                              &1.86   &1.83   &1.91\cite{Alonso_1997}\\
                  &d$_{Ni_o-O}$\footnotemark[1] [\AA]                 &2.00   &2.05  &2.12\cite{Alonso_1997}\\
                  &d$_{Ni_o-O}$\footnotemark[2] [\AA]                 &1.86   &1.88  &1.92\cite{Alonso_1997}\\
    \hline
     LaNiO$_{2}$  &d$_{Ni-Ni}$\footnotemark[1] [\AA]                  &3.89   &3.84   &3.96\cite{LNO2_PM} \\ 
           (NM)   &d$_{Ni-Ni}$\footnotemark[2] [\AA]                  &3.34   &3.32   &3.37\cite{LNO2_PM} \\ 
                &d$_{Ni-O}$\footnotemark[1] [\AA]                     &1.95   &1.92   &1.98\cite{LNO2_PM} \\ 
    \end{tabular}
    \footnotetext[1]{Along the $x-y$ plane.}
    \footnotetext[2]{Along the $z$ axis.}
    \end{ruledtabular}

\label{tbl:struc_thry_exp}
\end{table}

In Table$\:\:$\ref{tbl:total_eng}, we calculate the ground-state energies of LaNiO$_{3-x}$ using DFT and DFT+U. 
Different magnetic orderings including G-type AFM, FM and NM order are imposed during the relaxation calculations and subtract the resulting FM energy from the G-type AFM energy for each system.
In each structure, different magnetism can be obtained by converging the solutions from different initial configurations.
In LaNiO$_3$, the ground-state converges to the NM structure regardless of FM or AFM initial configurations in DFT while DFT+U predicts it to be FM. 
Also in LaNiO$_{2}$, DFT converges to the NM configuration while AFM is more stable in DFT+U. 
In LaNiO$_{2.5}$, DFT+U predicts the ground-state to be AFM consistently with the experiment (see Table$\:$\ref{tbl:magnetic}) while DFT converges to the FM ground state.
This implies that the correlation treated in DFT+U can be important to capture the ground-state magnetic configuration in structures with vacancies.
Both DFT and DFT+U give the FM order lower energy than AFM order in LaNiO$_{2.75}$, which is also consistent with experiments. 

In LaNiO$_{2.5}$, the spin-state ordering occurs as the Ni$_o$ ion in NiO$_6$ exhibits a high-spin state while the Ni$_{sp}$ ion in NiO$_4$ shows a low-spin state~\cite{Alonso_1997}. 
This spin-state ordering induced by the oxygen vacancy ordering is also consistent with our DFT+U calculation. We find the Ni$_O$ ion shows a high-spin state with the magnetic moment of 1.62$\mu_B$, which is much larger than the low-spin moment in Ni$_{sp}$ (0.16$\mu_B$). 
This spin-state ordering is also accompanied by the in-plane Ni-O bonding disproportionation in which the Ni$_o$-O bond length is much larger than the Ni$_{sp}$-O bond length by $\sim$0.22\AA$\:\:$ (see Table$\:$\ref{tbl:struc_thry_exp}). 
LaNiO$_{2.75}$ and LaNiO$_{2.25}$ with FM order also show the similar trends of the spin-state ordering in our calculations in that the Ni$_o$ ion has a larger moment than the Ni$_{sp}$ ion, as shown in Table~\ref{tbl:magnetic_DFT+U}.

\begin{table}[ht]
    \centering
    \caption{Total energy difference per formula unit [meV] between FM and AFM in LaNiO$_{3-x}$. } 
    \begin{tabular}{p{0.20\linewidth}|p{0.25\linewidth}|p{0.2\linewidth}}
    \hline\hline
      LaNiO$_{3-x}$ & DFT  & DFT+U \\ 
         & AFM-FM  & AFM-FM\\
      \hline
      LaNiO$_3$  & 0\footnotemark[1] & 341 \\ 
      LaNiO$_{2.75}$  &4 &   40  \\ 
      LaNiO$_{2.5}$   &10 &  -41 \\ 
      LaNiO$_{2.25}$  &2 & 68 \\ 
      LaNiO$_{2}$  &20\footnotemark[2] &-51 \\ 
    \hline\hline
    \end{tabular}
    \footnotetext[1]{Both FM and AFM converged to zero magnetic moment.}
    \footnotetext[2]{FM converged to zero magnetic moment.}
    \label{tbl:total_eng}
\end{table}

\begin{table}[ht]
    \centering
    \caption{Magnetic moments [$\mu B$] of LaNiO$_{3-x}$ computed using DFT+U.} 
    \begin{ruledtabular}
    \begin{tabular}{p{0.15\linewidth}p{0.2\linewidth}p{0.2\linewidth}}
      LaNiO$_{3-x}$ & Ni$_o$ & Ni$_{sp}$ \\ [1.0ex]
    \hline
      LaNiO$_{2.75}$[FM] & 1.04/1.47/1.04 & 0.11  \\ 
      LaNiO$_{2.5}$[AFM] & 1.62 & 0.16   \\ 
      LaNiO$_{2.25}$[FM] &1.60 & 0.8/0.8/0.8 \\ 
    \end{tabular}
    \end{ruledtabular}
\label{tbl:magnetic_DFT+U}
\end{table}

\subsection{Formation energies}
\label{sec:energetics}
The stability of the oxygen vacancy ordered structure can be determined by the formation energy calculation.
Here, we compute the vacancy formation energy per formula unit for LaNiO$_{3-x}$ structures as a function of the oxygen chemical potential related to the given oxygen pressure.
The formation energy can be given by~\cite{Geisler}
\begin{equation}
\label{eq:form_E}
E_{form}=E_{LaNiO_{3-x}}-E_{LaNiO_{3}}+x\cdot\frac{1}{2}E_{O_2}+x\cdot\mu_O
\end{equation}
where $E_{form}$ is the Gibbs formation energy, $x$ is the vacancy level, $E_{LaNiO_{3-x}}$ is the total energy of LaNiO$_{3-x}$, $E_{O_2}$ is the total energy of the $O_2$ molecule, and $\mu_O$ is the oxygen chemical potential depending on pressure and temperature. 
Here, we neglect the phonon and entropy contributions of LaNiO$_{3-x}$ to the Gibbs formation energy at finite temperatures. 
In experiments, the thermodynamic stability condition of oxygen vacancies in given materials can depend on the applied oxygen pressure $P$ and temperature $T$.
We assume that the oxygen molecule forms an ideal-gas-like reservoir during the experimental sample growth, therefore its chemical potential can be given by~\cite{For1,For2,For3,Geisler}
\begin{equation}
\mu_O(T,P)=\mu_O (T,P^{0})+\frac{1}{2}k_{B}T\ln\left(\frac{P}{P^{0}}\right)
\end{equation}
where $P^{0}$ is the ambient pressure. The $\mu_O (T,P^{0})$ values are taken from  Ref.$\:$\onlinecite{For1}. 
Typical experimental growth of LaNiO$_3$ on the SrTiO$_3$ substrate is carried out at around 920K and 10 Pa oxygen pressure~\cite{LaNiO3_PES_XAS}.

\begin{figure}
    \begin{subfigure}[b]{1\linewidth}
       \centering
       \includegraphics[width=\linewidth]{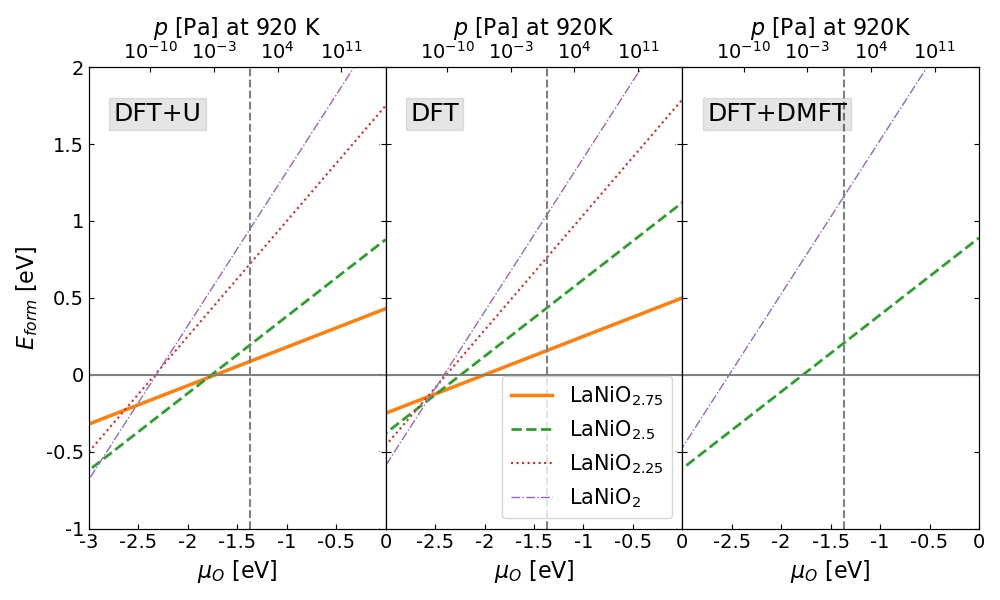}
    \end{subfigure}
\caption{Formation energies $E_{form}$ of LaNiO$_{3-x}$ as a function of the oxygen chemical potential $\mu_O$ (related to the oxygen pressure $P$) calculated using DFT+U (at $T=$0K), DFT (at $T=$0K) and DFT+DMFT (at $T=$920K). The vertical dashed line indicates the typical oxygen pressure used in experiment.}
\label{fig:formation}
\end{figure}

In Fig.$\:$\ref{fig:formation} we plot $E_{form}$ as a function of $\mu_O$ (related to the applied pressure) at 920K computed using DFT+U, DFT and DFT+DMFT. 
We compute the total energies of LaNiO$_{3-x}$ in Eq.$\:$\ref{eq:form_E} using first-principles (DFT, DFT+U, and DFT+DMFT) by adopting the same relaxed structures of LaNiO$_{3-x}$ as used in the spectra calculations.
We choose the magnetic order in DFT or DFT+U calculations resulting the lowest energy while DFT+DMFT calculations adopt the PM order for all LaNiO$_{3-x}$ structures at 920K.
Our results suggest that, for the oxygen-rich region, when the oxygen pressure higher than 55Pa (corresponding to $\mu_o$ $>$ -1.3 eV), LaNiO$_3$ is the most stable structure compared to other vacancy structures with the positive vacancy formation energies. However, as the oxygen pressure is lowered than 4.5Pa (corresponding to -2.1 eV $<$ $\mu_o$ $<$ -1.4 eV), we find that  LaNiO$_{2.5}$ becomes the most stable structure in the DFT+U calculation although different vacancy structures have very similar formation energies in DFT. 
DFT+DMFT also shows the similar energetics as a function of the oxygen pressure compared to DFT+U implying that correlation effects in DFT+DMFT or DFT+U treated beyond DFT can be important to capture the correct formation energy in oxygen vacancy structures.
Since DFT+U (DFT+DMFT) can capture the correct AFM (PM) order as well as the insulating state in LaNiO$_{2.5}$, the predicted vacancy ordered structure from first-principles can be indeed stable under the experimental growth condition of the lower oxygen pressure region.


\subsection{Correlated density of states} 
\label{sec:spectra}

\begin{figure}[b]
    \includegraphics[width=\linewidth]{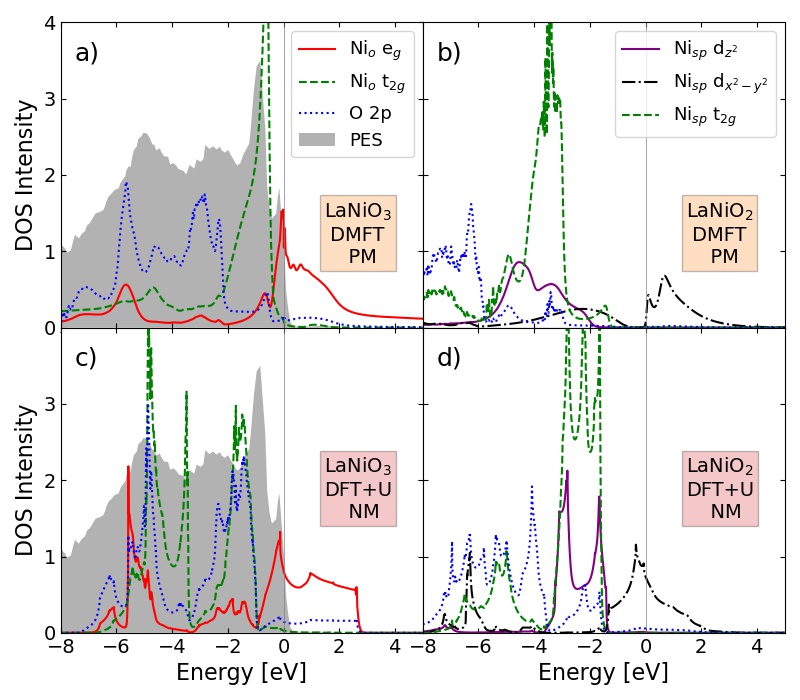}
\caption{The orbital-resolved DOS of LaNiO$_{3}$ (a,c) and LaNiO$_{2}$ (b,d) computed using DFT+DMFT (a,b) and DFT+U (c,d). The $e_g$ orbitals in LaNiO${_3}$ are degenerate. The shaded region is taken from the experimental PES measurement\cite{LaNiO3_PES_XAS}.}
\label{fig:LNO23}
\end{figure}

In this section, we study the correlated density of states (DOS) for the vacancy-ordered structure in LaNiO$_{2.5}$ as well as stoichiometric LaNiO${_3}$ and LaNiO${_2}$ using both DFT+DMFT and DFT+U to treat correlations beyond DFT.
First, we compare the DOS for stoichiometric LaNiO${_3}$ ($x$=0; Fig.$\:$\ref{fig:LNO23}a) and LaNiO${_2}$ ($x$=1; Fig.$\:$\ref{fig:LNO23}b) computed using DFT+DMFT. 
LaNiO$_3$ computed using DFT+DMFT shows the Fermi-liquid metal feature, which is consistent with the experimental measurement~\cite{PRB.79.115122}.
The overall peak positions computed in DFT+DMFT are also consistent with the experimental PES peak positions~\cite{LaNiO3_PES_XAS}.
Our orbital-resolved DFT+DMFT DOS reveals that the small bump below the Fermi energy has mostly the Ni $e_g$ character and the sharp peak near -0.7 eV is mostly contributed from the $t_{2g}$ character although O $2p$ orbitals are also mixed with these orbitals. O $2p$ peaks are distributed broadly below -2 eV. 
The unoccupied DOS also shows a broad $e_g$ peak with a strong mixture with O $2p$ spectra due to the covalent bonding between Ni and O ions.
As a result, the hole density per the O ion is 0.26 and the occupancy of the Ni 3$d$ orbital becomes $\sim$7.8 (See Table $\:$\ref{tbl:occ}).

The absence of apical oxygens in LaNiO$_2$ changes electronic structures significantly compared to LaNiO$_3$.
First, the apical oxygen vacancy breaks the symmetry of the Ni-O octahedron and split the on-site orbital energies within $e_g$ and $t_{2g}$ manifolds as the local geometry of the Ni-O bonding becomes a square-planar symmetry.
As a result, two $e_g$ orbitals in Ni$_{sp}$ become non-degenerate and the $d_{z^2}$ orbital is lower in energy than the $d_{x^2-y^2}$ orbital.
Second, the removal of the apical oxygen ions from the Ni-O octahedron means that two electrons are effectively donated to the remaining the Ni-O square plane. 
The donation of two electrons from the apical oxygen changes the oxidation state of Ni$^{3+}$ in LaNiO$_3$ to Ni$^{1+}$ in LaNiO$_2$. 
Due to this electron transfer, $d-$occupancy in Ni$_{sp}$ becomes close to 9.1 while the hole per O ion is also reduced to 0.06 (See Table $\:$\ref{tbl:occ}).
As a result, the $d_{z^2}$ orbital tends to be fully filled while the $d_{x^2-y^2}$ orbital is almost half-filled enhancing the electronic correlation effect.

Our DMFT DOS in LaNiO$_2$ is also consistent with the experimental XAS measurement~\cite{LNO_NNO_sc} showing the strong reduction of the unoccupied O 2$p$ spectra compared to the LaNiO$_3$ case. 
Our unoccupied O $2p$ DOS (blue dashed line in Fig.$\:$\ref{fig:LNO23}b) is also much reduced and the occupied O $2p$ peak in LaNiO$_2$ is located further below the Fermi energy due to the reduced hole density in the O ion and the decreased Ni-O hybridization compared to LaNiO${_3}$. 
Due to this much reduced Ni-O hybridization and the $d_{x^2-y^2}$ orbital occupancy close to the half-filling, the LaNiO$_2$ becomes close to the Mott insulating phase as reflected in the strong reduction of the $d_{x^2-y^2}$ spectra near the Fermi energy which is also consistent with the experimental XAS measurement~\cite{LNO_NNO_sc}. 
This incoherent metallic state due to the strong correlation effect is also reflected in the large scattering rate (the imaginary part of the self-energy at the zero frequency in Fig.$\:$\ref{fig:self-energy}b) and consistent with the poor conductivity measured experimentally in LaNiO$_2$~\cite{LNO2_Metallic}.
Our DFT band structure calculation reveals that some La 5$d$ bands are also crossing below the Fermi energy (see Supplemental Material) although we do not include this hybridization effect of the La 5$d$ orbital as we construct the Wannier functions for only Ni 3$d$ and O 2$p$ orbitals.
Although treating the effect of this La 5$d$ orbital on the DMFT correlation is beyond the scope of our paper, previous DMFT calculation in the similar NdNiO$_2$ material argues that the Nd 5$d$ band acts as a charge reservoir without significant hybridization with the Ni 3$d$ orbital~\cite{PhysRevX.10.021061}.

In Fig.$\:$\ref{fig:LNO23}c and \ref{fig:LNO23}d, we also plot the orbital-resolved DOS for both LaNiO${_3}$ and LaNiO${_2}$ obtained using DFT+U. 
We impose the NM spin order for both materials, consistently with the experiment as listed in Table$\:$\ref{tbl:magnetic}. 
In both materials, the DFT+U calculations also exhibit qualitatively similar features as the DFT+DMFT spectra.
In LaNiO$_3$, the $t_{2g}$ peak is located slightly below the experimental peak at -0.7eV and the it is somewhat strongly hybridized with O $p$ orbitals.
Consistently with DMFT, the unoccupied O 2$p$ DOS becomes much reduced in  LaNiO$_2$ and $d_{x^2-y^2}$ orbital is also nearly half filled.
However, the DFT+U spectra have still large DOS intensity at Fermi level and the O 2$p$ peak is relatively close to the Fermi energy compared to DMFT.

\begin{figure}[b]
    \begin{subfigure}[b]{1\linewidth}
       \centering
       \includegraphics[width=\linewidth]{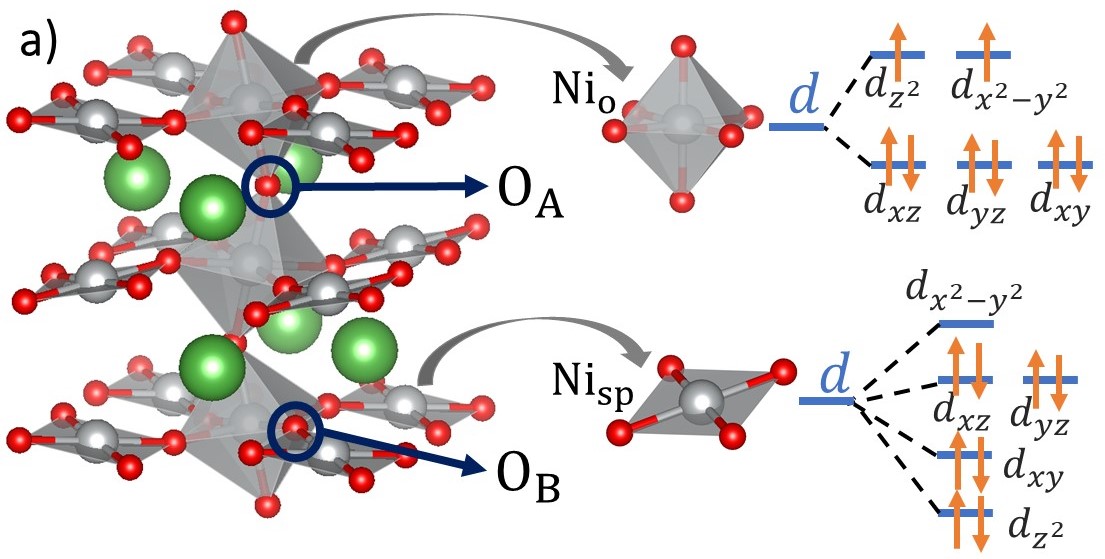}
    \end{subfigure}
    \begin{subfigure}[b]{\linewidth}
       \centering
    \includegraphics[width=\linewidth]{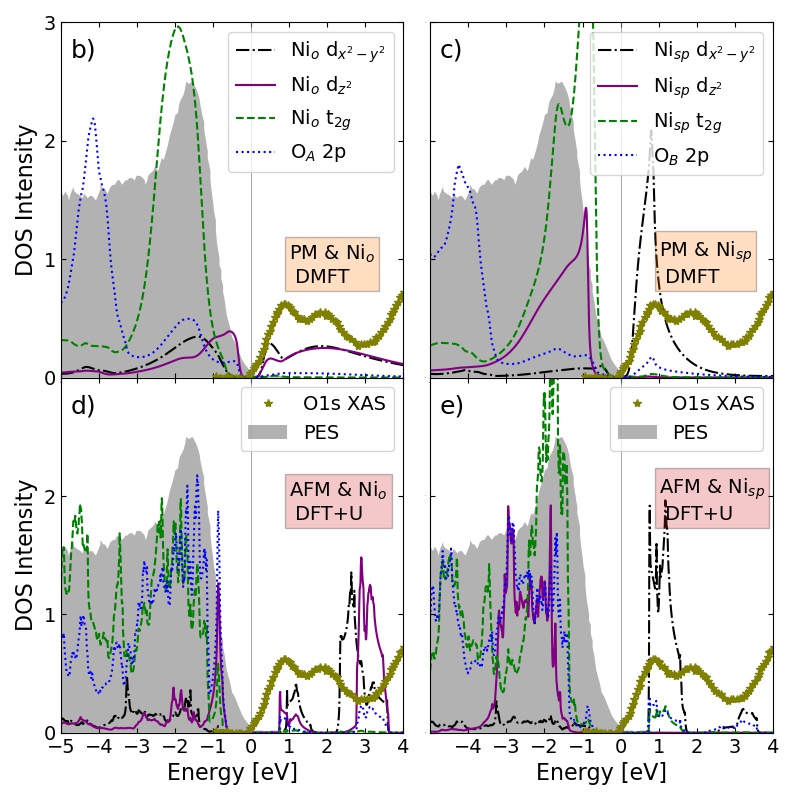}
    \end{subfigure}
\caption{(a) The crystal structure of NiO$_{2.5}$ showing two nonequivalent Ni ions with the distinct orbital energy level diagram. The site- and orbital-resolved DOS of LaNiO$_{2.5}$ obtained using DFT+DMFT (b and c) and DFT+U (d and e) computed for Ni$_o$ (b and d) and Ni$_{sp}$ (c and e). PES and XAS data are taken from experimental measurements~\cite{LaNiO3_PES_XAS}.}
\label{fig:LNO25}
\end{figure}

Now we turn to the LaNiO$_{2.5}$ case with the oxygen vacancy ordering.
We performed both DFT+DMFT and DFT+U calculations to study the correlated electronic structure beyond DFT treating the realistic oxygen vacancy structure (Fig.$\:$\ref{fig:LNO25}b-e). 
DFT+U was performed with the G-type AFM ordering (the lowest-energy magnetic structure; see Table$\:$\ref{tbl:total_eng}) and DFT+DMFT was performed with the paramagnetic spin symmetry. 
In this structure, the six-coordinated Ni-O octahedron (Ni$_o$) and the four-coordinated Ni-O square-plane (Ni$_{sp}$) are alternating in the $x-y$ plane as two apical oxygen ions are removed from the half of octahedra in LaNiO$_3$ (see Fig.$\:$\ref{fig:LNO25}a). Our structural relaxation calculation shows that the Ni-O octahedron is distorted as the in-plane Ni-O bond length is larger than the out-of-plane bond (see Table$\:$\ref{tbl:struc_thry_exp}). However, the orbital energy difference between two $e_g$ orbitals is only 0.04eV, which is much smaller than $e_g$-$t_{2g}$ splitting of 0.62eV. In Ni$_{sp}$, the energy splitting between $d_{z^2}$ and $d_{x^2-y^2}$ is as large as 0.8eV. 

As shown in Fig.$\:$\ref{fig:LNO25}b-e, both DFT+DMFT and DFT+U predict the LaNiO$_{2.5}$ to be insulating consistently with experiments. 
This is in sharp contrast with the DFT DOS predicting the ground state to be metallic as DFT underestimates correlations (see Supplemental Material).
DFT+DMFT DOS opens a spectral gap of 0.3 eV resulting in the PM insulating state while the band gap computed in DFT+U with AFM becomes much larger ($\sim$1.5 eV). 
Our DFT+DMFT calculation can also reveal that the two-peak structure of the unoccupied spectra measured in experiment originates from the two nonequivalent Ni ions in LaNiO$_{2.5}$.
The Ni$_o$ ion (Fig.$\:$\ref{fig:LNO25}b) develops a spectral Mott gap in the middle of the almost degenerate $e_g$ orbitals and the broad unoccupied peak near 2eV above the Fermi energy emerges as the upper Hubbard band due to the localized nature of Ni$_o$ $e_g$ orbitals induced by the oxygen vacancy.
In Ni$_{sp}$, the $d_{z^2}$ orbital spectra become occupied while the unoccupied DOS near 1eV is mostly contributed from the $d_{x^2-y^2}$ orbitals also hybridized with the in-plane O 2$p$ orbitals. 
Although all O ions are equivalent in LaNiO$_{3}$, two types of nonequivalent O ions exist in LaNiO$_{2.5}$, which are circled in Fig.$\:$\ref{fig:LNO25}a denoted as O$_A$ and O$_B$.
O$_A$ is the apical oxygen of the Ni$_o$ ion and O$_B$ is the in-plane oxygen located between Ni$_o$ and Ni$_{sp}$. 
The O1$s$ XAS spectra provide a relatively unperturbed probe of the unoccupied DOS since the O1$s$ core hole effect is negligible~\cite{PhysRevB.77.165127}.
However, it can probe not only O 2$p$ but also 3$d$ states in oxide materials due to strong $d-p$ hybridization~\cite{PhysRevResearch.2.033265}.
The O$_B$ 2$p$ unoccupied DOS is mostly distributed near 1eV as it is strongly hybridized with the Ni$_{sp}$ $d_{x^2-y^2}$ orbitals.
The O$_A$ 2$p$ unoccupied DOS is rather weakly hybridized with the Ni$_o$ ion showing much reduced spectra, but it is broadly distributed at higher energies than the O$_B$ spectra.
In DFT+U DOS (Fig.$\:$\ref{fig:LNO25}d and e), the qualitative features of DOS are similar as the DMFT DOS confirming that the unoccupied two-peak structure is originated from two nonequivalent Ni ions.
However, the DFT+U peak positions are higher in energy than the experimental peak positions.

\subsection{Site-selective Mott insulating state in LaNiO$_{2.5}$}
\label{sec:selfenergy}

\begin{figure}[b]
    \centering
       \includegraphics[width=\linewidth]{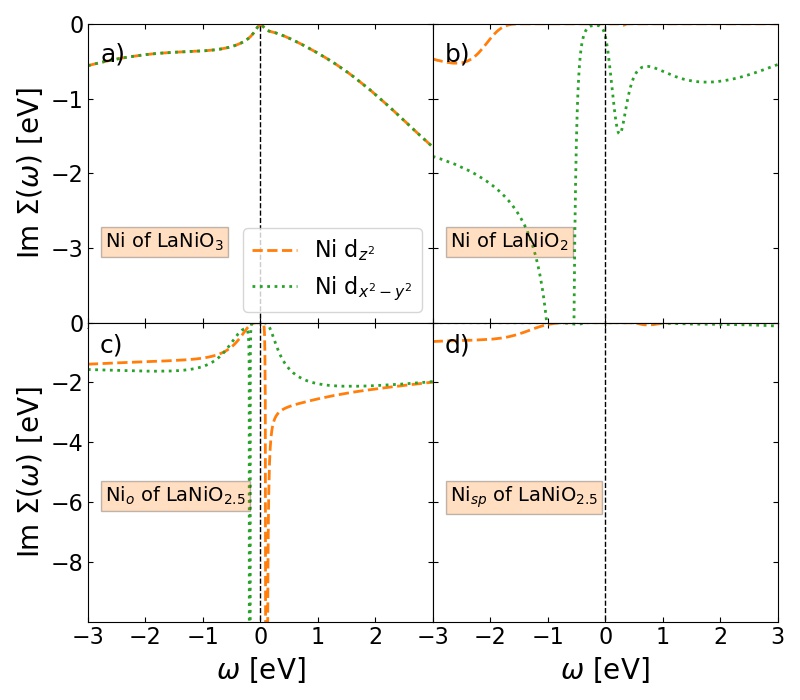}
\caption{Imaginary part of self-energies $Im\Sigma(\omega)$ computed for Ni d$_{z^{2}}$ and d$_{x^{2} - y^{2}}$ orbitals in (a) LaNiO$_{3}$, (b) LaNiO$_{2}$, and LaNiO$_{2.5}$ for two nonequivalent Ni ions, (c) Ni$_{o}$ and (d) Ni$_{sp}$ respectively.}
\label{fig:self-energy}
\end{figure}

Here, we show the imaginary part of the self-energy on the real energy axis, $Im\Sigma(\omega)$ of Ni $e_g$ orbitals computed using DMFT in Fig.$\:$\ref{fig:self-energy}. Two types of Ni ions in LaNiO$_{2.5}$ are plotted in Fig.$\:$\ref{fig:self-energy}c and Fig.$\:$\ref{fig:self-energy}d. LaNiO$_{3}$ (Fig.$\:$\ref{fig:self-energy}a) and LaNiO$_{2}$ (Fig.$\:$\ref{fig:self-energy}b) are also compared. The self-energies of $e_g$ orbitals in LaNiO$_{3}$ are degenerate and show the Fermi-liquid behavior ($Im \Sigma \sim \omega^2$) at the low frequency, consistently with the metallic state. In LaNiO$_2$, the $d_{x^2-y^2}$ orbital develop correlations as the self-energy develops a pole near the Fermi energy with the large scattering rate at $\omega=0$. In LaNiO$_{2.5}$, both Ni$_{o}$ e$_{g}$ orbitals have strong correlations at the Fermi energy as self energy curves have poles right near $\omega$=0 while Ni$_{sp}$ in LaNiO$_{2.5}$ exhibits a flat curve. Since Ni$_{o}$ e$_{g}$ orbitals in LaNiO$_{2.5}$ are also nearly half filled (see Fig.$\:$\ref{fig:LNO25}a energy diagram), this indicates a Mott type insulator in Ni$_{o}$ while the Ni$_{sp}$ ions with oxygen vacancies behave similarly as the band insulator as only one $e_{g}$ orbital is fully filled and the other $e_g$ orbital is almost unoccupied (see Fig.$\:$\ref{fig:LNO25}a energy diagram). This Mott insulating behavior occurs at the half of lattice sites selectivity (in this case for Ni$_{o}$ sites), therefore it can be understood as the ``site-selective'' Mott transition. 

This site-selective Mott insulating phase is also accompanied by the substantial Ni-O bond disproportionation of 0.22\AA$\:\:$ between Ni$_{o}$ and Ni$_{sp}$ as obtained in the DFT+U relaxation calculation (see Table \ref{tbl:struc_thry_exp}) since the Mott insulating site is expanded to reduce the hybridization between Ni and neighboring O ions.
In DFT+U, this mechanism occurs as the spin-state ordering since the Mott insulating site becomes a high-spin state while the band insulating site exhibits a low-spin state.
This site-selective Mott mechanism also occurs in other correlated materials including rare-earth nickelates with smaller rare-earth ions such as LuNiO$_3$~\cite{PhysRevLett.109.156402}, the spin-state ordered LaCoO$_3$~\cite{LaCoO3_SSMT}, Fe$_2$O$_3$ under the high pressure~\cite{PhysRevX.8.031059,Fe2O3_npj}, and doped manganese oxides~\cite{PhysRevB.92.115143}. Like the LaNiO$_3$ case with oxygen vacancies, the site-selective Mott insulating phase in other materials also accompanies some degrees of breathing-type lattice distortions as the Mott insulating site develops in an expanded octahedron and the band insulating site forms in a contracted octahedron due to strong hybridization of transition metal spins with surrounding O holes.

In DMFT calculation, one can also measure the correlation strength from the obtained quasi-particle residue $Z$ defined as: 
\begin{equation}
\label{eq:mass_renom}
Z=\left(\left.1-\frac{\partial \Sigma}{\partial \omega}\right|_{\omega=0}\right)^{-1}
\end{equation}
which gives the inverse of the effective mass renormalization factor ($m^*$/$m$=$Z^{-1}$). The Ni $e_g$ band of LaNiO${_3}$ has an effective mass factor of 1.7. For LaNiO$_2$, electronic correlation is further enhanced than LaNiO$_3$ and the quasi-particle band of the $d_{x^2-y^2}$ orbital is strongly renormalized with a factor of 9.4. As LaNiO$_{2.5}$ with oxygen vacancies becomes an insulator, the concept of the quasiparticle mass renormalization cannot be applied any more since quasiparticles do not exist in an insulator.

Oxygen vacancy also changes the Ni oxidation state and increases the electron occupation in the Ni-O manifold effectively as two electrons are donated from the removed oxygen ion. Table \ref{tbl:occ} shows the occupation number for Ni 3$d$ and O 2$p$ orbitals in each structure obtained from DFT+DMFT and DFT+U. 
In DFT+DMFT calculations, as the Ni oxidation state changes from Ni$^{3+}$ in LaNiO$_3$ to Ni$^{1+}$ in LaNiO$_2$, the $d-$orbital occupancy increases from 7.8 to 9.12 and the hole occupancy in the O ion decreases from 0.26 to 0.06. 
DFT+U also shows the increase of the average $d-$orbital occupancy from LaNiO$_3$ to LaNiO$_2$, however its value is larger than the DFT+DMFT result. 
In LaNiO$_{2.5}$, the average Ni oxidation state is Ni$^{2+}$. 
However, it is not clear whether the electron transfer due to oxygen vacancy will occur mostly to Ni$_{sp}$ resulting in charge ordering between Ni$^{3+}$ in the Ni$_{o}$ ion and Ni$^{1+}$ in the Ni$_{sp}$ ion or both Ni$_{o}$ and Ni$_{sp}$ ions will have the similar Ni$^{2+}$ configuration without charge ordering.
Our DFT+DMFT calculation shows that Ni$_{o}$ is close to Ni$^{2+}$ as the $e_g$ orbitals in Ni$_{o}$ are almost half filled ($\sim$2.13) and  the $e_g$ occupancy in Ni$_{sp}$ is slightly more occupied than the half-filling leaving some holes per the O site ($\sim$0.15). 
This hole state in O is rather strongly hybridized with the Ni$_{sp}$ $e_g$ orbital as shown in the previous DOS result.
Therefore, the site-selective Mott transition in LaNiO$_{2.5}$ also leads to small charge ordering between Ni$_{o}$ and Ni$_{sp}$ ions.
DFT+U also shows the similar charge ordering in LaNiO$_{2.5}$.

\begin{table}[ht]
    \centering 
    \caption{Occupations of Ni 3$d$ and O 2$p$ orbitals in LaNiO$_{3-x}$ obtained from DFT+DMFT and DFT+U.}
    \begin{tabular}{p{0.32\linewidth}|p{0.15\linewidth}p{0.15\linewidth}p{0.1\linewidth}}
    \hline\hline
     DFT+DMFT& Ni$_{o}$ & Ni$_{sp}$ & O$_{avg}$\\
     \hline
      LaNiO$_3$(PM) & 7.79 & N/A & 5.74 \\ 
      LaNiO$_{2.5}$(PM) & 8.15 &8.57  &5.86  \\ 
      LaNiO$_{2}$(PM) & N/A & 9.12& 5.94\\ 
     \hline\hline
     DFT+U&&&\\
     \hline
      LaNiO$_3$(NM) & 8.23 & N/A & 5.59\\ 
      LaNiO$_{2.5}$(AFM) & 8.27 &8.73  &5.80  \\ 
      LaNiO$_{2}$(NM) & N/A & 9.30& 5.85\\ 
     \hline\hline
    \end{tabular}
    \label{tbl:occ}
\end{table}

\subsection{Rigid band shift approximation in LaNiO$_{3-x}$}
\label{sec:rigidband}

Our DFT+DMFT calculation in LaNiO$_{2.5}$ shows that the paramagnetic insulating phase in LaNiO$_{2.5}$ originates from the change of Ni oxidation states as well as the oxygen vacancy ordering structure which induces both the local symmetry change of Ni ions and different hybridization of Ni ions with surrounding O ions. 
To investigate the effect of the oxygen vacancy ordering structure on electronic correlations, we apply the rigid band shift approximation to LaNiO$_{3}$ within DFT+DMFT to impose the effect of the Ni oxidation state change alone. 
In this approximation, we use the same Wannier band structure obtained from LaNiO$_{3}$ at different vacancy levels while the effect of different Ni oxidation states is adopted by shifting the Fermi level to modify the total number of electrons within DMFT calculations accordingly.

\begin{figure}[h]
    \begin{subfigure}[b]{0.85\linewidth}
       \centering
       \includegraphics[width=\linewidth]{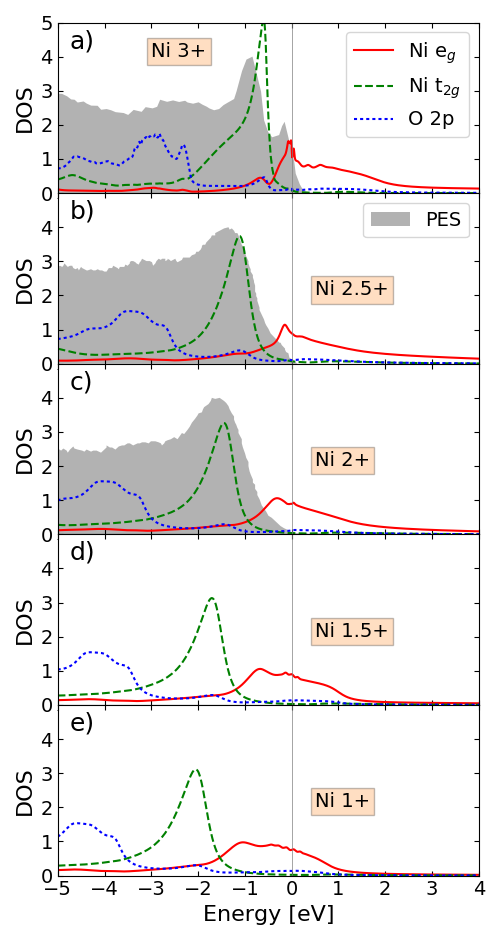}
    \end{subfigure}

    \begin{subfigure}[b]{\linewidth}
       \centering
    \includegraphics[width=\linewidth]{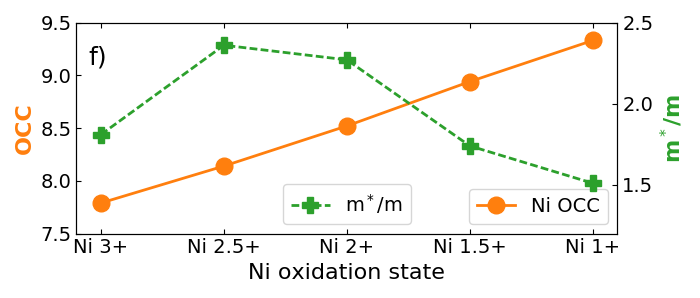}
    \end{subfigure}

\caption{(a)-(e) DFT+DMFT DOS of LaNiO$_{3-x}$ systems at different Ni oxidation states computed using the rigid band shift approximation. (f) The $d$ occupancy and the mass renormalization factor as a function of different Ni oxidation states.}
\label{fig:rigid-band}
\end{figure}

Fig.$\:$\ref{fig:rigid-band}a-e shows the DOS of different Ni oxidation states due to the change of vacancy level $x$, namely Ni$^{3+}$ for $x$=0 and Ni$^{1+}$ for $x$=1. 
As the oxidation number changes from Ni$^{3+}$ to Ni$^{1+}$, O $p$ and Ni $t_{2g}$ states move further below the Fermi energy while keep the shape mostly unchanged.
This trend is also consistent with the experiment data depicted as the shaded region.
The Ni $e_{g}$ states near the Fermi energy also do not change significantly as the Fermi level shifts higher in energy for the smaller oxidation state.
The Ni$^{2+}$ state corresponding to  LaNiO$_{2.5}$ still exhibits the metallic state without developing a Mott state although the $e_g$ occupancy becomes close to the half-filling. 
In Fig.\ref{fig:rigid-band}f, we plot the mass renormalization factor $m^*/m$  and the corresponding Ni $d$ occupancies as a function of Ni oxidation states. While the $d$ occupancies change linearly, the effective mass becomes maximum only when the $d$ occupancy becomes close to 8.0 meaning the half-filled $e_g$ occupancy although the mass renormalization remains in range of 1.5-2.5.
Therefore, the strong correlation effect occurring in the oxygen vacancy structure cannot be captured by the Ni oxidation change alone within the rigid band shift approximation.

\section{Conclusion}
\label{sec:conclusion}
In conclusion, we performed first principles calculations in LaNiO$_3$ with oxygen vacancies (LaNiO$_{3-x}$ with $x$=0, 0.25, 0.5, 0.75, and 1).
Experimentally, the metal-to-insulator transition occurs as the oxygen vacancy level $x$ approaches to 0.5 and the ground state is AFM (AFM becomes PM at higher temperature).
We find that the vacancy ordering structure of alternating NiO$_6$ octahedra and NiO$_4$ square planes (with apical oxygen vacancies) becomes thermodynamically stable in LaNiO$_{2.5}$ when the oxygen pressure is lowered than the typical growth condition of LaNiO$_3$.
DFT+U converges to the correct AFM magnetic order in LaNiO$_{2.5}$ while DFT favors FM implying that the electronic correlation effect can be important for oxygen vacancy structures.
Our DFT+DMFT calculation in LaNiO$_{2.5}$ shows that the PM state is insulating and the nature of this insulating state is the site-selective Mott phase in which the octahedral Ni ion develops a Mott state with strong electron correlations and the square-planar Ni ion becomes a band insulator with the negligible self-energy.
We also explain the nature of the two-peak structure in unoccupied spectra measured in O1s XAS experiments of LaNiO$_{3-x}$.
The lower energy peak in XAS originates from the square-planar Ni state strongly hybridized with O ions while the higher energy peak is resulted from the broad spectra of the localized $e_g$ orbitals in strongly correlated octahedral Ni ion.
LaNiO$_{2}$ with the complete apical oxygen vacancies becomes also strongly correlated as it reduces the Ni-O hybridization and lifts the degeneracy between $e_g$ orbitals, as a result, it becomes close to the Mott state  although it is still metallic.

The change of Ni oxidation states alone within the rigid band shift approximation cannot capture the strongly correlated insulating phase occurring in the oxygen vacancy structure of LaNiO$_3$ implying that it is important to treat realistic oxygen vacancy structures of materials using first-principles to account for electronic correlation effects induced by oxygen vacancies.
Moreover, the Hubbard $U$ values of Ni ions can be site-dependent in oxygen-vacancy structures since chemical environments of Ni ions can be varied due to oxygen vacancies. In principle, site-dependent $U$ values can be computed from first-principles, and their effects on electronic structure of materials with oxygen vacancies can be an interesting topic for future studies.

\section*{Acknowledgement}
X. Liao and H. Park are supported by the Materials Sciences and Engineering Division, Basic Energy Sciences, Office of Science, US DOE. The part of this work related to the vacancy formation energy calculation is supported by ACS-PRF grant 60617. V. Singh is supported by the NSF SI2-SSE Grant 1740112. We gratefully acknowledge the computing resources provided on Bebop, a high-performance computing cluster operated by the Laboratory Computing Resource Center at Argonne National Laboratory.

\bibliography{ref.bib}

\begin{thebibliography}{63}%
\makeatletter
\providecommand \@ifxundefined [1]{%
 \@ifx{#1\undefined}
}%
\providecommand \@ifnum [1]{%
 \ifnum #1\expandafter \@firstoftwo
 \else \expandafter \@secondoftwo
 \fi
}%
\providecommand \@ifx [1]{%
 \ifx #1\expandafter \@firstoftwo
 \else \expandafter \@secondoftwo
 \fi
}%
\providecommand \natexlab [1]{#1}%
\providecommand \enquote  [1]{``#1''}%
\providecommand \bibnamefont  [1]{#1}%
\providecommand \bibfnamefont [1]{#1}%
\providecommand \citenamefont [1]{#1}%
\providecommand \href@noop [0]{\@secondoftwo}%
\providecommand \href [0]{\begingroup \@sanitize@url \@href}%
\providecommand \@href[1]{\@@startlink{#1}\@@href}%
\providecommand \@@href[1]{\endgroup#1\@@endlink}%
\providecommand \@sanitize@url [0]{\catcode `\\12\catcode `\$12\catcode
  `\&12\catcode `\#12\catcode `\^12\catcode `\_12\catcode `\%12\relax}%
\providecommand \@@startlink[1]{}%
\providecommand \@@endlink[0]{}%
\providecommand \url  [0]{\begingroup\@sanitize@url \@url }%
\providecommand \@url [1]{\endgroup\@href {#1}{\urlprefix }}%
\providecommand \urlprefix  [0]{URL }%
\providecommand \Eprint [0]{\href }%
\providecommand \doibase [0]{https://doi.org/}%
\providecommand \selectlanguage [0]{\@gobble}%
\providecommand \bibinfo  [0]{\@secondoftwo}%
\providecommand \bibfield  [0]{\@secondoftwo}%
\providecommand \translation [1]{[#1]}%
\providecommand \BibitemOpen [0]{}%
\providecommand \bibitemStop [0]{}%
\providecommand \bibitemNoStop [0]{.\EOS\space}%
\providecommand \EOS [0]{\spacefactor3000\relax}%
\providecommand \BibitemShut  [1]{\csname bibitem#1\endcsname}%
\let\auto@bib@innerbib\@empty
\bibitem [{\citenamefont {Catalan}(2008)}]{rev_RNO}%
  \BibitemOpen
  \bibfield  {author} {\bibinfo {author} {\bibfnamefont {G.}~\bibnamefont
  {Catalan}},\ }\href {https://doi.org/10.1080/01411590801992463} {\bibfield
  {journal} {\bibinfo  {journal} {Phase Transitions}\ }\textbf {\bibinfo
  {volume} {81}},\ \bibinfo {pages} {729} (\bibinfo {year} {2008})},\ \Eprint
  {https://arxiv.org/abs/https://doi.org/10.1080/01411590801992463}
  {https://doi.org/10.1080/01411590801992463} \BibitemShut {NoStop}%
\bibitem [{\citenamefont {Zubko}\ \emph {et~al.}(2011)\citenamefont {Zubko},
  \citenamefont {Gariglio}, \citenamefont {Gabay}, \citenamefont {Ghosez},\
  and\ \citenamefont {Triscone}}]{rev_RNO_2}%
  \BibitemOpen
  \bibfield  {author} {\bibinfo {author} {\bibfnamefont {P.}~\bibnamefont
  {Zubko}}, \bibinfo {author} {\bibfnamefont {S.}~\bibnamefont {Gariglio}},
  \bibinfo {author} {\bibfnamefont {M.}~\bibnamefont {Gabay}}, \bibinfo
  {author} {\bibfnamefont {P.}~\bibnamefont {Ghosez}},\ and\ \bibinfo {author}
  {\bibfnamefont {J.-M.}\ \bibnamefont {Triscone}},\ }\href
  {https://doi.org/10.1146/annurev-conmatphys-062910-140445} {\bibfield
  {journal} {\bibinfo  {journal} {Annual Review of Condensed Matter Physics}\
  }\textbf {\bibinfo {volume} {2}},\ \bibinfo {pages} {141} (\bibinfo {year}
  {2011})},\ \Eprint
  {https://arxiv.org/abs/https://doi.org/10.1146/annurev-conmatphys-062910-140445}
  {https://doi.org/10.1146/annurev-conmatphys-062910-140445} \BibitemShut
  {NoStop}%
\bibitem [{\citenamefont {Alonso}\ \emph {et~al.}(2000)\citenamefont {Alonso},
  \citenamefont {Mart\'{\i}nez-Lope}, \citenamefont {Casais}, \citenamefont
  {Garc\'{\i}a-Mu\~noz},\ and\ \citenamefont
  {Fern\'andez-D\'{\i}az}}]{RNO_charge_order}%
  \BibitemOpen
  \bibfield  {author} {\bibinfo {author} {\bibfnamefont {J.~A.}\ \bibnamefont
  {Alonso}}, \bibinfo {author} {\bibfnamefont {M.~J.}\ \bibnamefont
  {Mart\'{\i}nez-Lope}}, \bibinfo {author} {\bibfnamefont {M.~T.}\ \bibnamefont
  {Casais}}, \bibinfo {author} {\bibfnamefont {J.~L.}\ \bibnamefont
  {Garc\'{\i}a-Mu\~noz}},\ and\ \bibinfo {author} {\bibfnamefont {M.~T.}\
  \bibnamefont {Fern\'andez-D\'{\i}az}},\ }\href
  {https://doi.org/10.1103/PhysRevB.61.1756} {\bibfield  {journal} {\bibinfo
  {journal} {Phys. Rev. B}\ }\textbf {\bibinfo {volume} {61}},\ \bibinfo
  {pages} {1756} (\bibinfo {year} {2000})}\BibitemShut {NoStop}%
\bibitem [{\citenamefont {Staub}\ \emph {et~al.}(2002)\citenamefont {Staub},
  \citenamefont {Meijer}, \citenamefont {Fauth}, \citenamefont {Allenspach},
  \citenamefont {Bednorz}, \citenamefont {Karpinski}, \citenamefont {Kazakov},
  \citenamefont {Paolasini},\ and\ \citenamefont
  {d'Acapito}}]{NNO_charge_order}%
  \BibitemOpen
  \bibfield  {author} {\bibinfo {author} {\bibfnamefont {U.}~\bibnamefont
  {Staub}}, \bibinfo {author} {\bibfnamefont {G.~I.}\ \bibnamefont {Meijer}},
  \bibinfo {author} {\bibfnamefont {F.}~\bibnamefont {Fauth}}, \bibinfo
  {author} {\bibfnamefont {R.}~\bibnamefont {Allenspach}}, \bibinfo {author}
  {\bibfnamefont {J.~G.}\ \bibnamefont {Bednorz}}, \bibinfo {author}
  {\bibfnamefont {J.}~\bibnamefont {Karpinski}}, \bibinfo {author}
  {\bibfnamefont {S.~M.}\ \bibnamefont {Kazakov}}, \bibinfo {author}
  {\bibfnamefont {L.}~\bibnamefont {Paolasini}},\ and\ \bibinfo {author}
  {\bibfnamefont {F.}~\bibnamefont {d'Acapito}},\ }\href
  {https://doi.org/10.1103/PhysRevLett.88.126402} {\bibfield  {journal}
  {\bibinfo  {journal} {Phys. Rev. Lett.}\ }\textbf {\bibinfo {volume} {88}},\
  \bibinfo {pages} {126402} (\bibinfo {year} {2002})}\BibitemShut {NoStop}%
\bibitem [{\citenamefont {Zhou}\ \emph {et~al.}(2000)\citenamefont {Zhou},
  \citenamefont {Goodenough}, \citenamefont {Dabrowski}, \citenamefont
  {Klamut},\ and\ \citenamefont {Bukowski}}]{RNO_mag_trans}%
  \BibitemOpen
  \bibfield  {author} {\bibinfo {author} {\bibfnamefont {J.-S.}\ \bibnamefont
  {Zhou}}, \bibinfo {author} {\bibfnamefont {J.~B.}\ \bibnamefont
  {Goodenough}}, \bibinfo {author} {\bibfnamefont {B.}~\bibnamefont
  {Dabrowski}}, \bibinfo {author} {\bibfnamefont {P.~W.}\ \bibnamefont
  {Klamut}},\ and\ \bibinfo {author} {\bibfnamefont {Z.}~\bibnamefont
  {Bukowski}},\ }\href {https://doi.org/10.1103/PhysRevLett.84.526} {\bibfield
  {journal} {\bibinfo  {journal} {Phys. Rev. Lett.}\ }\textbf {\bibinfo
  {volume} {84}},\ \bibinfo {pages} {526} (\bibinfo {year} {2000})}\BibitemShut
  {NoStop}%
\bibitem [{\citenamefont {Park}\ \emph {et~al.}(2012)\citenamefont {Park},
  \citenamefont {Millis},\ and\ \citenamefont
  {Marianetti}}]{PhysRevLett.109.156402}%
  \BibitemOpen
  \bibfield  {author} {\bibinfo {author} {\bibfnamefont {H.}~\bibnamefont
  {Park}}, \bibinfo {author} {\bibfnamefont {A.~J.}\ \bibnamefont {Millis}},\
  and\ \bibinfo {author} {\bibfnamefont {C.~A.}\ \bibnamefont {Marianetti}},\
  }\href {https://doi.org/10.1103/PhysRevLett.109.156402} {\bibfield  {journal}
  {\bibinfo  {journal} {Phys. Rev. Lett.}\ }\textbf {\bibinfo {volume} {109}},\
  \bibinfo {pages} {156402} (\bibinfo {year} {2012})}\BibitemShut {NoStop}%
\bibitem [{\citenamefont {Catalano}\ \emph {et~al.}(2018)\citenamefont
  {Catalano}, \citenamefont {Gibert}, \citenamefont {Fowlie}, \citenamefont
  {{\'{I}}{\~{n}}iguez}, \citenamefont {Triscone},\ and\ \citenamefont
  {Kreisel}}]{RNO_MIT_rev}%
  \BibitemOpen
  \bibfield  {author} {\bibinfo {author} {\bibfnamefont {S.}~\bibnamefont
  {Catalano}}, \bibinfo {author} {\bibfnamefont {M.}~\bibnamefont {Gibert}},
  \bibinfo {author} {\bibfnamefont {J.}~\bibnamefont {Fowlie}}, \bibinfo
  {author} {\bibfnamefont {J.}~\bibnamefont {{\'{I}}{\~{n}}iguez}}, \bibinfo
  {author} {\bibfnamefont {J.-M.}\ \bibnamefont {Triscone}},\ and\ \bibinfo
  {author} {\bibfnamefont {J.}~\bibnamefont {Kreisel}},\ }\href
  {https://doi.org/10.1088/1361-6633/aaa37a} {\bibfield  {journal} {\bibinfo
  {journal} {Reports on Progress in Physics}\ }\textbf {\bibinfo {volume}
  {81}},\ \bibinfo {pages} {046501} (\bibinfo {year} {2018})}\BibitemShut
  {NoStop}%
\bibitem [{\citenamefont {Gunkel}\ \emph {et~al.}(2020)\citenamefont {Gunkel},
  \citenamefont {Christensen}, \citenamefont {Chen},\ and\ \citenamefont
  {Pryds}}]{doi:10.1063/1.5143309}%
  \BibitemOpen
  \bibfield  {author} {\bibinfo {author} {\bibfnamefont {F.}~\bibnamefont
  {Gunkel}}, \bibinfo {author} {\bibfnamefont {D.~V.}\ \bibnamefont
  {Christensen}}, \bibinfo {author} {\bibfnamefont {Y.~Z.}\ \bibnamefont
  {Chen}},\ and\ \bibinfo {author} {\bibfnamefont {N.}~\bibnamefont {Pryds}},\
  }\href {https://doi.org/10.1063/1.5143309} {\bibfield  {journal} {\bibinfo
  {journal} {Applied Physics Letters}\ }\textbf {\bibinfo {volume} {116}},\
  \bibinfo {pages} {120505} (\bibinfo {year} {2020})}\BibitemShut {NoStop}%
\bibitem [{\citenamefont {Eguchi}\ \emph {et~al.}(2009)\citenamefont {Eguchi},
  \citenamefont {Chainani}, \citenamefont {Taguchi}, \citenamefont {Matsunami},
  \citenamefont {Ishida}, \citenamefont {Horiba}, \citenamefont {Senba},
  \citenamefont {Ohashi},\ and\ \citenamefont {Shin}}]{PRB.79.115122}%
  \BibitemOpen
  \bibfield  {author} {\bibinfo {author} {\bibfnamefont {R.}~\bibnamefont
  {Eguchi}}, \bibinfo {author} {\bibfnamefont {A.}~\bibnamefont {Chainani}},
  \bibinfo {author} {\bibfnamefont {M.}~\bibnamefont {Taguchi}}, \bibinfo
  {author} {\bibfnamefont {M.}~\bibnamefont {Matsunami}}, \bibinfo {author}
  {\bibfnamefont {Y.}~\bibnamefont {Ishida}}, \bibinfo {author} {\bibfnamefont
  {K.}~\bibnamefont {Horiba}}, \bibinfo {author} {\bibfnamefont
  {Y.}~\bibnamefont {Senba}}, \bibinfo {author} {\bibfnamefont
  {H.}~\bibnamefont {Ohashi}},\ and\ \bibinfo {author} {\bibfnamefont
  {S.}~\bibnamefont {Shin}},\ }\href
  {https://doi.org/10.1103/PhysRevB.79.115122} {\bibfield  {journal} {\bibinfo
  {journal} {Phys. Rev. B}\ }\textbf {\bibinfo {volume} {79}},\ \bibinfo
  {pages} {115122} (\bibinfo {year} {2009})}\BibitemShut {NoStop}%
\bibitem [{\citenamefont {Stewart}\ \emph {et~al.}(2011)\citenamefont
  {Stewart}, \citenamefont {Yee}, \citenamefont {Liu}, \citenamefont {Kareev},
  \citenamefont {Smith}, \citenamefont {Chapler}, \citenamefont {Varela},
  \citenamefont {Ryan}, \citenamefont {Haule}, \citenamefont {Chakhalian},\
  and\ \citenamefont {Basov}}]{PRB.83.075125}%
  \BibitemOpen
  \bibfield  {author} {\bibinfo {author} {\bibfnamefont {M.~K.}\ \bibnamefont
  {Stewart}}, \bibinfo {author} {\bibfnamefont {C.-H.}\ \bibnamefont {Yee}},
  \bibinfo {author} {\bibfnamefont {J.}~\bibnamefont {Liu}}, \bibinfo {author}
  {\bibfnamefont {M.}~\bibnamefont {Kareev}}, \bibinfo {author} {\bibfnamefont
  {R.~K.}\ \bibnamefont {Smith}}, \bibinfo {author} {\bibfnamefont {B.~C.}\
  \bibnamefont {Chapler}}, \bibinfo {author} {\bibfnamefont {M.}~\bibnamefont
  {Varela}}, \bibinfo {author} {\bibfnamefont {P.~J.}\ \bibnamefont {Ryan}},
  \bibinfo {author} {\bibfnamefont {K.}~\bibnamefont {Haule}}, \bibinfo
  {author} {\bibfnamefont {J.}~\bibnamefont {Chakhalian}},\ and\ \bibinfo
  {author} {\bibfnamefont {D.~N.}\ \bibnamefont {Basov}},\ }\href
  {https://doi.org/10.1103/PhysRevB.83.075125} {\bibfield  {journal} {\bibinfo
  {journal} {Phys. Rev. B}\ }\textbf {\bibinfo {volume} {83}},\ \bibinfo
  {pages} {075125} (\bibinfo {year} {2011})}\BibitemShut {NoStop}%
\bibitem [{\citenamefont {Ouellette}\ \emph {et~al.}(2010)\citenamefont
  {Ouellette}, \citenamefont {Lee}, \citenamefont {Son}, \citenamefont
  {Stemmer}, \citenamefont {Balents}, \citenamefont {Millis},\ and\
  \citenamefont {Allen}}]{PRB.82.165112}%
  \BibitemOpen
  \bibfield  {author} {\bibinfo {author} {\bibfnamefont {D.~G.}\ \bibnamefont
  {Ouellette}}, \bibinfo {author} {\bibfnamefont {S.~B.}\ \bibnamefont {Lee}},
  \bibinfo {author} {\bibfnamefont {J.}~\bibnamefont {Son}}, \bibinfo {author}
  {\bibfnamefont {S.}~\bibnamefont {Stemmer}}, \bibinfo {author} {\bibfnamefont
  {L.}~\bibnamefont {Balents}}, \bibinfo {author} {\bibfnamefont {A.~J.}\
  \bibnamefont {Millis}},\ and\ \bibinfo {author} {\bibfnamefont {S.~J.}\
  \bibnamefont {Allen}},\ }\href {https://doi.org/10.1103/PhysRevB.82.165112}
  {\bibfield  {journal} {\bibinfo  {journal} {Phys. Rev. B}\ }\textbf {\bibinfo
  {volume} {82}},\ \bibinfo {pages} {165112} (\bibinfo {year}
  {2010})}\BibitemShut {NoStop}%
\bibitem [{\citenamefont {Moriga}\ \emph {et~al.}(1995)\citenamefont {Moriga},
  \citenamefont {Usaka}, \citenamefont {Nakabayashi}, \citenamefont {Kinouchi},
  \citenamefont {Kikkawa},\ and\ \citenamefont
  {Kanamaru}}]{MORIGA1995252_LNO2.6}%
  \BibitemOpen
  \bibfield  {author} {\bibinfo {author} {\bibfnamefont {T.}~\bibnamefont
  {Moriga}}, \bibinfo {author} {\bibfnamefont {O.}~\bibnamefont {Usaka}},
  \bibinfo {author} {\bibfnamefont {I.}~\bibnamefont {Nakabayashi}}, \bibinfo
  {author} {\bibfnamefont {T.}~\bibnamefont {Kinouchi}}, \bibinfo {author}
  {\bibfnamefont {S.}~\bibnamefont {Kikkawa}},\ and\ \bibinfo {author}
  {\bibfnamefont {F.}~\bibnamefont {Kanamaru}},\ }\href
  {https://doi.org/https://doi.org/10.1016/0167-2738(95)00070-M} {\bibfield
  {journal} {\bibinfo  {journal} {Solid State Ionics}\ }\textbf {\bibinfo
  {volume} {79}},\ \bibinfo {pages} {252 } (\bibinfo {year} {1995})},\ \bibinfo
  {note} {proceedings of the 20th Commemorative Symposium on Solid State Ionics
  in Japan}\BibitemShut {NoStop}%
\bibitem [{\citenamefont {S\'anchez}\ \emph {et~al.}(1996)\citenamefont
  {S\'anchez}, \citenamefont {Causa}, \citenamefont {Caneiro}, \citenamefont
  {Butera}, \citenamefont {Vallet-Reg\'{\i}}, \citenamefont {Sayagu\'es},
  \citenamefont {Gonz\'alez-Calbet}, \citenamefont {Garc\'{\i}a-Sanz},\ and\
  \citenamefont {Rivas}}]{Sanches_MIT}%
  \BibitemOpen
  \bibfield  {author} {\bibinfo {author} {\bibfnamefont {R.~D.}\ \bibnamefont
  {S\'anchez}}, \bibinfo {author} {\bibfnamefont {M.~T.}\ \bibnamefont
  {Causa}}, \bibinfo {author} {\bibfnamefont {A.}~\bibnamefont {Caneiro}},
  \bibinfo {author} {\bibfnamefont {A.}~\bibnamefont {Butera}}, \bibinfo
  {author} {\bibfnamefont {M.}~\bibnamefont {Vallet-Reg\'{\i}}}, \bibinfo
  {author} {\bibfnamefont {M.~J.}\ \bibnamefont {Sayagu\'es}}, \bibinfo
  {author} {\bibfnamefont {J.}~\bibnamefont {Gonz\'alez-Calbet}}, \bibinfo
  {author} {\bibfnamefont {F.}~\bibnamefont {Garc\'{\i}a-Sanz}},\ and\ \bibinfo
  {author} {\bibfnamefont {J.}~\bibnamefont {Rivas}},\ }\href
  {https://doi.org/10.1103/PhysRevB.54.16574} {\bibfield  {journal} {\bibinfo
  {journal} {Phys. Rev. B}\ }\textbf {\bibinfo {volume} {54}},\ \bibinfo
  {pages} {16574} (\bibinfo {year} {1996})}\BibitemShut {NoStop}%
\bibitem [{\citenamefont {Kaneko}\ \emph {et~al.}(2009)\citenamefont {Kaneko},
  \citenamefont {Yamagishi}, \citenamefont {Tsukada}, \citenamefont {Manabe},\
  and\ \citenamefont {Naito}}]{LNO2_Metallic}%
  \BibitemOpen
  \bibfield  {author} {\bibinfo {author} {\bibfnamefont {D.}~\bibnamefont
  {Kaneko}}, \bibinfo {author} {\bibfnamefont {K.}~\bibnamefont {Yamagishi}},
  \bibinfo {author} {\bibfnamefont {A.}~\bibnamefont {Tsukada}}, \bibinfo
  {author} {\bibfnamefont {T.}~\bibnamefont {Manabe}},\ and\ \bibinfo {author}
  {\bibfnamefont {M.}~\bibnamefont {Naito}},\ }\href
  {https://doi.org/https://doi.org/10.1016/j.physc.2009.05.104} {\bibfield
  {journal} {\bibinfo  {journal} {Physica C: Superconductivity}\ }\textbf
  {\bibinfo {volume} {469}},\ \bibinfo {pages} {936 } (\bibinfo {year}
  {2009})},\ \bibinfo {note} {proceedings of the 21st International Symposium
  on Superconductivity (ISS 2008)}\BibitemShut {NoStop}%
\bibitem [{\citenamefont {Li}\ \emph {et~al.}(2019)\citenamefont {Li},
  \citenamefont {Lee}, \citenamefont {Wang}, \citenamefont {Osada},
  \citenamefont {Crossley}, \citenamefont {Lee}, \citenamefont {Cui},
  \citenamefont {Hikita},\ and\ \citenamefont {Hwang}}]{nature_NdNiO2_sc}%
  \BibitemOpen
  \bibfield  {author} {\bibinfo {author} {\bibfnamefont {D.}~\bibnamefont
  {Li}}, \bibinfo {author} {\bibfnamefont {K.}~\bibnamefont {Lee}}, \bibinfo
  {author} {\bibfnamefont {B.}~\bibnamefont {Wang}}, \bibinfo {author}
  {\bibfnamefont {M.}~\bibnamefont {Osada}}, \bibinfo {author} {\bibfnamefont
  {S.}~\bibnamefont {Crossley}}, \bibinfo {author} {\bibfnamefont {H.~R.}\
  \bibnamefont {Lee}}, \bibinfo {author} {\bibfnamefont {Y.}~\bibnamefont
  {Cui}}, \bibinfo {author} {\bibfnamefont {Y.}~\bibnamefont {Hikita}},\ and\
  \bibinfo {author} {\bibfnamefont {H.~Y.}\ \bibnamefont {Hwang}},\ }\href
  {https://doi.org/10.1038/s41586-019-1496-5} {\bibfield  {journal} {\bibinfo
  {journal} {Nature}\ }\textbf {\bibinfo {volume} {572}},\ \bibinfo {pages}
  {624} (\bibinfo {year} {2019})}\BibitemShut {NoStop}%
\bibitem [{\citenamefont {Osada}\ \emph {et~al.}(2020)\citenamefont {Osada},
  \citenamefont {Wang}, \citenamefont {Goodge}, \citenamefont {Lee},
  \citenamefont {Yoon}, \citenamefont {Sakuma}, \citenamefont {Li},
  \citenamefont {Miura}, \citenamefont {Kourkoutis},\ and\ \citenamefont
  {Hwang}}]{PrNiO2}%
  \BibitemOpen
  \bibfield  {author} {\bibinfo {author} {\bibfnamefont {M.}~\bibnamefont
  {Osada}}, \bibinfo {author} {\bibfnamefont {B.~Y.}\ \bibnamefont {Wang}},
  \bibinfo {author} {\bibfnamefont {B.~H.}\ \bibnamefont {Goodge}}, \bibinfo
  {author} {\bibfnamefont {K.}~\bibnamefont {Lee}}, \bibinfo {author}
  {\bibfnamefont {H.}~\bibnamefont {Yoon}}, \bibinfo {author} {\bibfnamefont
  {K.}~\bibnamefont {Sakuma}}, \bibinfo {author} {\bibfnamefont
  {D.}~\bibnamefont {Li}}, \bibinfo {author} {\bibfnamefont {M.}~\bibnamefont
  {Miura}}, \bibinfo {author} {\bibfnamefont {L.~F.}\ \bibnamefont
  {Kourkoutis}},\ and\ \bibinfo {author} {\bibfnamefont {H.~Y.}\ \bibnamefont
  {Hwang}},\ }\href {https://doi.org/10.1021/acs.nanolett.0c01392} {\bibfield
  {journal} {\bibinfo  {journal} {Nano Letters}\ }\textbf {\bibinfo {volume}
  {20}},\ \bibinfo {pages} {5735} (\bibinfo {year} {2020})}\BibitemShut
  {NoStop}%
\bibitem [{\citenamefont {Guo}\ \emph {et~al.}(2018)\citenamefont {Guo},
  \citenamefont {Li}, \citenamefont {Zhao}, \citenamefont {Hu}, \citenamefont
  {Chang}, \citenamefont {Kuo}, \citenamefont {Schmidt}, \citenamefont
  {Piovano}, \citenamefont {Pi}, \citenamefont {Sobolev}, \citenamefont
  {Khomskii}, \citenamefont {Tjeng},\ and\ \citenamefont {Komarek}}]{LNO3_afm}%
  \BibitemOpen
  \bibfield  {author} {\bibinfo {author} {\bibfnamefont {H.}~\bibnamefont
  {Guo}}, \bibinfo {author} {\bibfnamefont {Z.~W.}\ \bibnamefont {Li}},
  \bibinfo {author} {\bibfnamefont {L.}~\bibnamefont {Zhao}}, \bibinfo {author}
  {\bibfnamefont {Z.}~\bibnamefont {Hu}}, \bibinfo {author} {\bibfnamefont
  {C.~F.}\ \bibnamefont {Chang}}, \bibinfo {author} {\bibfnamefont {C.-Y.}\
  \bibnamefont {Kuo}}, \bibinfo {author} {\bibfnamefont {W.}~\bibnamefont
  {Schmidt}}, \bibinfo {author} {\bibfnamefont {A.}~\bibnamefont {Piovano}},
  \bibinfo {author} {\bibfnamefont {T.~W.}\ \bibnamefont {Pi}}, \bibinfo
  {author} {\bibfnamefont {O.}~\bibnamefont {Sobolev}}, \bibinfo {author}
  {\bibfnamefont {D.~I.}\ \bibnamefont {Khomskii}}, \bibinfo {author}
  {\bibfnamefont {L.~H.}\ \bibnamefont {Tjeng}},\ and\ \bibinfo {author}
  {\bibfnamefont {A.~C.}\ \bibnamefont {Komarek}},\ }\bibfield  {journal}
  {\bibinfo  {journal} {Nature Communications}\ }\textbf {\bibinfo {volume}
  {9}},\ \href {https://doi.org/10.1038/s41467-017-02524-x}
  {10.1038/s41467-017-02524-x} (\bibinfo {year} {2018})\BibitemShut {NoStop}%
\bibitem [{\citenamefont {Zhang}\ \emph {et~al.}(2017)\citenamefont {Zhang},
  \citenamefont {Zheng}, \citenamefont {Ren},\ and\ \citenamefont
  {Mitchell}}]{LNO3_PM2}%
  \BibitemOpen
  \bibfield  {author} {\bibinfo {author} {\bibfnamefont {J.}~\bibnamefont
  {Zhang}}, \bibinfo {author} {\bibfnamefont {H.}~\bibnamefont {Zheng}},
  \bibinfo {author} {\bibfnamefont {Y.}~\bibnamefont {Ren}},\ and\ \bibinfo
  {author} {\bibfnamefont {J.~F.}\ \bibnamefont {Mitchell}},\ }\href
  {https://doi.org/10.1021/acs.cgd.7b00205} {\bibfield  {journal} {\bibinfo
  {journal} {Crystal Growth \& Design}\ }\textbf {\bibinfo {volume} {17}},\
  \bibinfo {pages} {2730} (\bibinfo {year} {2017})},\ \Eprint
  {https://arxiv.org/abs/https://doi.org/10.1021/acs.cgd.7b00205}
  {https://doi.org/10.1021/acs.cgd.7b00205} \BibitemShut {NoStop}%
\bibitem [{\citenamefont {Wang}\ \emph {et~al.}(2018)\citenamefont {Wang},
  \citenamefont {Rosenkranz}, \citenamefont {Rui}, \citenamefont {Zhang},
  \citenamefont {Ye}, \citenamefont {Zheng}, \citenamefont {Klie},
  \citenamefont {Mitchell},\ and\ \citenamefont {Phelan}}]{LNO_FM_AFM_2018}%
  \BibitemOpen
  \bibfield  {author} {\bibinfo {author} {\bibfnamefont {B.-X.}\ \bibnamefont
  {Wang}}, \bibinfo {author} {\bibfnamefont {S.}~\bibnamefont {Rosenkranz}},
  \bibinfo {author} {\bibfnamefont {X.}~\bibnamefont {Rui}}, \bibinfo {author}
  {\bibfnamefont {J.}~\bibnamefont {Zhang}}, \bibinfo {author} {\bibfnamefont
  {F.}~\bibnamefont {Ye}}, \bibinfo {author} {\bibfnamefont {H.}~\bibnamefont
  {Zheng}}, \bibinfo {author} {\bibfnamefont {R.~F.}\ \bibnamefont {Klie}},
  \bibinfo {author} {\bibfnamefont {J.~F.}\ \bibnamefont {Mitchell}},\ and\
  \bibinfo {author} {\bibfnamefont {D.}~\bibnamefont {Phelan}},\ }\href
  {https://doi.org/10.1103/PhysRevMaterials.2.064404} {\bibfield  {journal}
  {\bibinfo  {journal} {Phys. Rev. Materials}\ }\textbf {\bibinfo {volume}
  {2}},\ \bibinfo {pages} {064404} (\bibinfo {year} {2018})}\BibitemShut
  {NoStop}%
\bibitem [{\citenamefont {Hayward}\ \emph {et~al.}(1999)\citenamefont
  {Hayward}, \citenamefont {Green}, \citenamefont {Rosseinsky},\ and\
  \citenamefont {Sloan}}]{LNO2_PM}%
  \BibitemOpen
  \bibfield  {author} {\bibinfo {author} {\bibfnamefont {M.~A.}\ \bibnamefont
  {Hayward}}, \bibinfo {author} {\bibfnamefont {M.~A.}\ \bibnamefont {Green}},
  \bibinfo {author} {\bibfnamefont {M.~J.}\ \bibnamefont {Rosseinsky}},\ and\
  \bibinfo {author} {\bibfnamefont {J.}~\bibnamefont {Sloan}},\ }\href
  {https://doi.org/10.1021/ja991573i} {\bibfield  {journal} {\bibinfo
  {journal} {Journal of the American Chemical Society}\ }\textbf {\bibinfo
  {volume} {121}},\ \bibinfo {pages} {8843} (\bibinfo {year} {1999})},\ \Eprint
  {https://arxiv.org/abs/https://doi.org/10.1021/ja991573i}
  {https://doi.org/10.1021/ja991573i} \BibitemShut {NoStop}%
\bibitem [{\citenamefont {Abbate}\ \emph {et~al.}(2002)\citenamefont {Abbate},
  \citenamefont {Zampieri}, \citenamefont {Prado}, \citenamefont {Caneiro},
  \citenamefont {Gonzalez-Calbet},\ and\ \citenamefont {Vallet-Regi}}]{XAS}%
  \BibitemOpen
  \bibfield  {author} {\bibinfo {author} {\bibfnamefont {M.}~\bibnamefont
  {Abbate}}, \bibinfo {author} {\bibfnamefont {G.}~\bibnamefont {Zampieri}},
  \bibinfo {author} {\bibfnamefont {F.}~\bibnamefont {Prado}}, \bibinfo
  {author} {\bibfnamefont {A.}~\bibnamefont {Caneiro}}, \bibinfo {author}
  {\bibfnamefont {J.~M.}\ \bibnamefont {Gonzalez-Calbet}},\ and\ \bibinfo
  {author} {\bibfnamefont {M.}~\bibnamefont {Vallet-Regi}},\ }\href
  {https://doi.org/10.1103/PhysRevB.65.155101} {\bibfield  {journal} {\bibinfo
  {journal} {Phys. Rev. B}\ }\textbf {\bibinfo {volume} {65}},\ \bibinfo
  {pages} {155101} (\bibinfo {year} {2002})}\BibitemShut {NoStop}%
\bibitem [{\citenamefont {Horiba}\ \emph {et~al.}(2007)\citenamefont {Horiba},
  \citenamefont {Eguchi}, \citenamefont {Taguchi}, \citenamefont {Chainani},
  \citenamefont {Kikkawa}, \citenamefont {Senba}, \citenamefont {Ohashi},\ and\
  \citenamefont {Shin}}]{LaNiO3_PES_XAS}%
  \BibitemOpen
  \bibfield  {author} {\bibinfo {author} {\bibfnamefont {K.}~\bibnamefont
  {Horiba}}, \bibinfo {author} {\bibfnamefont {R.}~\bibnamefont {Eguchi}},
  \bibinfo {author} {\bibfnamefont {M.}~\bibnamefont {Taguchi}}, \bibinfo
  {author} {\bibfnamefont {A.}~\bibnamefont {Chainani}}, \bibinfo {author}
  {\bibfnamefont {A.}~\bibnamefont {Kikkawa}}, \bibinfo {author} {\bibfnamefont
  {Y.}~\bibnamefont {Senba}}, \bibinfo {author} {\bibfnamefont
  {H.}~\bibnamefont {Ohashi}},\ and\ \bibinfo {author} {\bibfnamefont
  {S.}~\bibnamefont {Shin}},\ }\href
  {https://doi.org/10.1103/PhysRevB.76.155104} {\bibfield  {journal} {\bibinfo
  {journal} {Phys. Rev. B}\ }\textbf {\bibinfo {volume} {76}},\ \bibinfo
  {pages} {155104} (\bibinfo {year} {2007})}\BibitemShut {NoStop}%
\bibitem [{\citenamefont {Golalikhani}\ \emph {et~al.}(2018)\citenamefont
  {Golalikhani}, \citenamefont {Lei}, \citenamefont {Chandrasena},
  \citenamefont {Kasaei}, \citenamefont {Park}, \citenamefont {Bai},
  \citenamefont {Orgiani}, \citenamefont {Ciston}, \citenamefont {Sterbinsky},
  \citenamefont {Arena}, \citenamefont {Shafer}, \citenamefont {Arenholz},
  \citenamefont {Davidson}, \citenamefont {Millis}, \citenamefont {Gray},\ and\
  \citenamefont {Xi}}]{Natcomm18}%
  \BibitemOpen
  \bibfield  {author} {\bibinfo {author} {\bibfnamefont {M.}~\bibnamefont
  {Golalikhani}}, \bibinfo {author} {\bibfnamefont {Q.}~\bibnamefont {Lei}},
  \bibinfo {author} {\bibfnamefont {R.~U.}\ \bibnamefont {Chandrasena}},
  \bibinfo {author} {\bibfnamefont {L.}~\bibnamefont {Kasaei}}, \bibinfo
  {author} {\bibfnamefont {H.}~\bibnamefont {Park}}, \bibinfo {author}
  {\bibfnamefont {J.}~\bibnamefont {Bai}}, \bibinfo {author} {\bibfnamefont
  {P.}~\bibnamefont {Orgiani}}, \bibinfo {author} {\bibfnamefont
  {J.}~\bibnamefont {Ciston}}, \bibinfo {author} {\bibfnamefont {G.~E.}\
  \bibnamefont {Sterbinsky}}, \bibinfo {author} {\bibfnamefont {D.~A.}\
  \bibnamefont {Arena}}, \bibinfo {author} {\bibfnamefont {P.}~\bibnamefont
  {Shafer}}, \bibinfo {author} {\bibfnamefont {E.}~\bibnamefont {Arenholz}},
  \bibinfo {author} {\bibfnamefont {B.~A.}\ \bibnamefont {Davidson}}, \bibinfo
  {author} {\bibfnamefont {A.~J.}\ \bibnamefont {Millis}}, \bibinfo {author}
  {\bibfnamefont {A.~X.}\ \bibnamefont {Gray}},\ and\ \bibinfo {author}
  {\bibfnamefont {X.~X.}\ \bibnamefont {Xi}},\ }\href@noop {} {\bibfield
  {journal} {\bibinfo  {journal} {Nature Communications}\ }\textbf {\bibinfo
  {volume} {9}},\ \bibinfo {pages} {2206} (\bibinfo {year} {2018})}\BibitemShut
  {NoStop}%
\bibitem [{\citenamefont {Misra}\ and\ \citenamefont {Kundu}(2016)}]{EPJB16}%
  \BibitemOpen
  \bibfield  {author} {\bibinfo {author} {\bibfnamefont {D.}~\bibnamefont
  {Misra}}\ and\ \bibinfo {author} {\bibfnamefont {T.~K.}\ \bibnamefont
  {Kundu}},\ }\href@noop {} {\bibfield  {journal} {\bibinfo  {journal} {The
  European Physical Journal B}\ }\textbf {\bibinfo {volume} {89}},\ \bibinfo
  {pages} {4} (\bibinfo {year} {2016})}\BibitemShut {NoStop}%
\bibitem [{\citenamefont {Malashevich}\ and\ \citenamefont
  {Ismail-Beigi}(2015)}]{DFT_study}%
  \BibitemOpen
  \bibfield  {author} {\bibinfo {author} {\bibfnamefont {A.}~\bibnamefont
  {Malashevich}}\ and\ \bibinfo {author} {\bibfnamefont {S.}~\bibnamefont
  {Ismail-Beigi}},\ }\href {https://doi.org/10.1103/PhysRevB.92.144102}
  {\bibfield  {journal} {\bibinfo  {journal} {Phys. Rev. B}\ }\textbf {\bibinfo
  {volume} {92}},\ \bibinfo {pages} {144102} (\bibinfo {year}
  {2015})}\BibitemShut {NoStop}%
\bibitem [{\citenamefont {Kotiuga}\ \emph {et~al.}(2019)\citenamefont
  {Kotiuga}, \citenamefont {Zhang}, \citenamefont {Li}, \citenamefont
  {Rodolakis}, \citenamefont {Zhou}, \citenamefont {Sutarto}, \citenamefont
  {He}, \citenamefont {Wang}, \citenamefont {Sun}, \citenamefont {Wang},
  \citenamefont {Aghamiri}, \citenamefont {Hancock}, \citenamefont {Rokhinson},
  \citenamefont {Landau}, \citenamefont {Abate}, \citenamefont {Freeland},
  \citenamefont {Comin}, \citenamefont {Ramanathan},\ and\ \citenamefont
  {Rabe}}]{Kotiuga21992}%
  \BibitemOpen
  \bibfield  {author} {\bibinfo {author} {\bibfnamefont {M.}~\bibnamefont
  {Kotiuga}}, \bibinfo {author} {\bibfnamefont {Z.}~\bibnamefont {Zhang}},
  \bibinfo {author} {\bibfnamefont {J.}~\bibnamefont {Li}}, \bibinfo {author}
  {\bibfnamefont {F.}~\bibnamefont {Rodolakis}}, \bibinfo {author}
  {\bibfnamefont {H.}~\bibnamefont {Zhou}}, \bibinfo {author} {\bibfnamefont
  {R.}~\bibnamefont {Sutarto}}, \bibinfo {author} {\bibfnamefont
  {F.}~\bibnamefont {He}}, \bibinfo {author} {\bibfnamefont {Q.}~\bibnamefont
  {Wang}}, \bibinfo {author} {\bibfnamefont {Y.}~\bibnamefont {Sun}}, \bibinfo
  {author} {\bibfnamefont {Y.}~\bibnamefont {Wang}}, \bibinfo {author}
  {\bibfnamefont {N.~A.}\ \bibnamefont {Aghamiri}}, \bibinfo {author}
  {\bibfnamefont {S.~B.}\ \bibnamefont {Hancock}}, \bibinfo {author}
  {\bibfnamefont {L.~P.}\ \bibnamefont {Rokhinson}}, \bibinfo {author}
  {\bibfnamefont {D.~P.}\ \bibnamefont {Landau}}, \bibinfo {author}
  {\bibfnamefont {Y.}~\bibnamefont {Abate}}, \bibinfo {author} {\bibfnamefont
  {J.~W.}\ \bibnamefont {Freeland}}, \bibinfo {author} {\bibfnamefont
  {R.}~\bibnamefont {Comin}}, \bibinfo {author} {\bibfnamefont
  {S.}~\bibnamefont {Ramanathan}},\ and\ \bibinfo {author} {\bibfnamefont
  {K.~M.}\ \bibnamefont {Rabe}},\ }\href
  {https://doi.org/10.1073/pnas.1910490116} {\bibfield  {journal} {\bibinfo
  {journal} {Proceedings of the National Academy of Sciences}\ }\textbf
  {\bibinfo {volume} {116}},\ \bibinfo {pages} {21992} (\bibinfo {year}
  {2019})},\ \Eprint
  {https://arxiv.org/abs/https://www.pnas.org/content/116/44/21992.full.pdf}
  {https://www.pnas.org/content/116/44/21992.full.pdf} \BibitemShut {NoStop}%
\bibitem [{\citenamefont {Subedi}(2018)}]{10.21468/SciPostPhys.5.3.020}%
  \BibitemOpen
  \bibfield  {author} {\bibinfo {author} {\bibfnamefont {A.}~\bibnamefont
  {Subedi}},\ }\href {https://doi.org/10.21468/SciPostPhys.5.3.020} {\bibfield
  {journal} {\bibinfo  {journal} {SciPost Phys.}\ }\textbf {\bibinfo {volume}
  {5}},\ \bibinfo {pages} {20} (\bibinfo {year} {2018})}\BibitemShut {NoStop}%
\bibitem [{\citenamefont {Bandyopadhyay}\ \emph {et~al.}(2020)\citenamefont
  {Bandyopadhyay}, \citenamefont {Adhikary}, \citenamefont {Das}, \citenamefont
  {Dasgupta},\ and\ \citenamefont {Saha-Dasgupta}}]{NNO_PNO_LNO_sc}%
  \BibitemOpen
  \bibfield  {author} {\bibinfo {author} {\bibfnamefont {S.}~\bibnamefont
  {Bandyopadhyay}}, \bibinfo {author} {\bibfnamefont {P.}~\bibnamefont
  {Adhikary}}, \bibinfo {author} {\bibfnamefont {T.}~\bibnamefont {Das}},
  \bibinfo {author} {\bibfnamefont {I.}~\bibnamefont {Dasgupta}},\ and\
  \bibinfo {author} {\bibfnamefont {T.}~\bibnamefont {Saha-Dasgupta}},\ }\href
  {https://doi.org/10.1103/PhysRevB.102.220502} {\bibfield  {journal} {\bibinfo
   {journal} {Phys. Rev. B}\ }\textbf {\bibinfo {volume} {102}},\ \bibinfo
  {pages} {220502(R)} (\bibinfo {year} {2020})}\BibitemShut {NoStop}%
\bibitem [{\citenamefont {Botana}\ and\ \citenamefont
  {Norman}(2020)}]{LNO_CCO_sc}%
  \BibitemOpen
  \bibfield  {author} {\bibinfo {author} {\bibfnamefont {A.~S.}\ \bibnamefont
  {Botana}}\ and\ \bibinfo {author} {\bibfnamefont {M.~R.}\ \bibnamefont
  {Norman}},\ }\href {https://doi.org/10.1103/PhysRevX.10.011024} {\bibfield
  {journal} {\bibinfo  {journal} {Phys. Rev. X}\ }\textbf {\bibinfo {volume}
  {10}},\ \bibinfo {pages} {011024} (\bibinfo {year} {2020})}\BibitemShut
  {NoStop}%
\bibitem [{\citenamefont {Karp}\ \emph {et~al.}(2020)\citenamefont {Karp},
  \citenamefont {Botana}, \citenamefont {Norman}, \citenamefont {Park},
  \citenamefont {Zingl},\ and\ \citenamefont {Millis}}]{PhysRevX.10.021061}%
  \BibitemOpen
  \bibfield  {author} {\bibinfo {author} {\bibfnamefont {J.}~\bibnamefont
  {Karp}}, \bibinfo {author} {\bibfnamefont {A.~S.}\ \bibnamefont {Botana}},
  \bibinfo {author} {\bibfnamefont {M.~R.}\ \bibnamefont {Norman}}, \bibinfo
  {author} {\bibfnamefont {H.}~\bibnamefont {Park}}, \bibinfo {author}
  {\bibfnamefont {M.}~\bibnamefont {Zingl}},\ and\ \bibinfo {author}
  {\bibfnamefont {A.}~\bibnamefont {Millis}},\ }\href
  {https://doi.org/10.1103/PhysRevX.10.021061} {\bibfield  {journal} {\bibinfo
  {journal} {Phys. Rev. X}\ }\textbf {\bibinfo {volume} {10}},\ \bibinfo
  {pages} {021061} (\bibinfo {year} {2020})}\BibitemShut {NoStop}%
\bibitem [{\citenamefont {Ryee}\ \emph {et~al.}(2020)\citenamefont {Ryee},
  \citenamefont {Yoon}, \citenamefont {Kim}, \citenamefont {Jeong},\ and\
  \citenamefont {Han}}]{ryee2019induced}%
  \BibitemOpen
  \bibfield  {author} {\bibinfo {author} {\bibfnamefont {S.}~\bibnamefont
  {Ryee}}, \bibinfo {author} {\bibfnamefont {H.}~\bibnamefont {Yoon}}, \bibinfo
  {author} {\bibfnamefont {T.~J.}\ \bibnamefont {Kim}}, \bibinfo {author}
  {\bibfnamefont {M.~Y.}\ \bibnamefont {Jeong}},\ and\ \bibinfo {author}
  {\bibfnamefont {M.~J.}\ \bibnamefont {Han}},\ }\href
  {https://doi.org/10.1103/PhysRevB.101.064513} {\bibfield  {journal} {\bibinfo
   {journal} {Phys. Rev. B}\ }\textbf {\bibinfo {volume} {101}},\ \bibinfo
  {pages} {064513} (\bibinfo {year} {2020})}\BibitemShut {NoStop}%
\bibitem [{\citenamefont {Werner}\ and\ \citenamefont
  {Hoshino}(2020)}]{werner2019nickelate}%
  \BibitemOpen
  \bibfield  {author} {\bibinfo {author} {\bibfnamefont {P.}~\bibnamefont
  {Werner}}\ and\ \bibinfo {author} {\bibfnamefont {S.}~\bibnamefont
  {Hoshino}},\ }\href {https://doi.org/10.1103/PhysRevB.101.041104} {\bibfield
  {journal} {\bibinfo  {journal} {Phys. Rev. B}\ }\textbf {\bibinfo {volume}
  {101}},\ \bibinfo {pages} {041104(R)} (\bibinfo {year} {2020})}\BibitemShut
  {NoStop}%
\bibitem [{\citenamefont {Lechermann}(2020)}]{lechermann2019late}%
  \BibitemOpen
  \bibfield  {author} {\bibinfo {author} {\bibfnamefont {F.}~\bibnamefont
  {Lechermann}},\ }\href {https://doi.org/10.1103/PhysRevB.101.081110}
  {\bibfield  {journal} {\bibinfo  {journal} {Phys. Rev. B}\ }\textbf {\bibinfo
  {volume} {101}},\ \bibinfo {pages} {081110(R)} (\bibinfo {year}
  {2020})}\BibitemShut {NoStop}%
\bibitem [{\citenamefont {Wang}\ \emph {et~al.}(2020)\citenamefont {Wang},
  \citenamefont {Kang}, \citenamefont {Miao},\ and\ \citenamefont
  {Kotliar}}]{PhysRevB.102.161118}%
  \BibitemOpen
  \bibfield  {author} {\bibinfo {author} {\bibfnamefont {Y.}~\bibnamefont
  {Wang}}, \bibinfo {author} {\bibfnamefont {C.-J.}\ \bibnamefont {Kang}},
  \bibinfo {author} {\bibfnamefont {H.}~\bibnamefont {Miao}},\ and\ \bibinfo
  {author} {\bibfnamefont {G.}~\bibnamefont {Kotliar}},\ }\href
  {https://doi.org/10.1103/PhysRevB.102.161118} {\bibfield  {journal} {\bibinfo
   {journal} {Phys. Rev. B}\ }\textbf {\bibinfo {volume} {102}},\ \bibinfo
  {pages} {161118(R)} (\bibinfo {year} {2020})}\BibitemShut {NoStop}%
\bibitem [{\citenamefont {Kresse}\ and\ \citenamefont
  {Furthm\"uller}(1996)}]{vasp1}%
  \BibitemOpen
  \bibfield  {author} {\bibinfo {author} {\bibfnamefont {G.}~\bibnamefont
  {Kresse}}\ and\ \bibinfo {author} {\bibfnamefont {J.}~\bibnamefont
  {Furthm\"uller}},\ }\href {https://doi.org/10.1103/PhysRevB.54.11169}
  {\bibfield  {journal} {\bibinfo  {journal} {Phys. Rev. B}\ }\textbf {\bibinfo
  {volume} {54}},\ \bibinfo {pages} {11169} (\bibinfo {year}
  {1996})}\BibitemShut {NoStop}%
\bibitem [{\citenamefont {Kresse}\ and\ \citenamefont {Joubert}(1999)}]{vasp2}%
  \BibitemOpen
  \bibfield  {author} {\bibinfo {author} {\bibfnamefont {G.}~\bibnamefont
  {Kresse}}\ and\ \bibinfo {author} {\bibfnamefont {D.}~\bibnamefont
  {Joubert}},\ }\href {https://doi.org/10.1103/PhysRevB.59.1758} {\bibfield
  {journal} {\bibinfo  {journal} {Phys. Rev. B}\ }\textbf {\bibinfo {volume}
  {59}},\ \bibinfo {pages} {1758} (\bibinfo {year} {1999})}\BibitemShut
  {NoStop}%
\bibitem [{\citenamefont {Perdew}\ \emph {et~al.}(2008)\citenamefont {Perdew},
  \citenamefont {Ruzsinszky}, \citenamefont {Csonka}, \citenamefont {Vydrov},
  \citenamefont {Scuseria}, \citenamefont {Constantin}, \citenamefont {Zhou},\
  and\ \citenamefont {Burke}}]{PBEsol}%
  \BibitemOpen
  \bibfield  {author} {\bibinfo {author} {\bibfnamefont {J.~P.}\ \bibnamefont
  {Perdew}}, \bibinfo {author} {\bibfnamefont {A.}~\bibnamefont {Ruzsinszky}},
  \bibinfo {author} {\bibfnamefont {G.~I.}\ \bibnamefont {Csonka}}, \bibinfo
  {author} {\bibfnamefont {O.~A.}\ \bibnamefont {Vydrov}}, \bibinfo {author}
  {\bibfnamefont {G.~E.}\ \bibnamefont {Scuseria}}, \bibinfo {author}
  {\bibfnamefont {L.~A.}\ \bibnamefont {Constantin}}, \bibinfo {author}
  {\bibfnamefont {X.}~\bibnamefont {Zhou}},\ and\ \bibinfo {author}
  {\bibfnamefont {K.}~\bibnamefont {Burke}},\ }\href
  {https://doi.org/10.1103/PhysRevLett.100.136406} {\bibfield  {journal}
  {\bibinfo  {journal} {Phys. Rev. Lett.}\ }\textbf {\bibinfo {volume} {100}},\
  \bibinfo {pages} {136406} (\bibinfo {year} {2008})}\BibitemShut {NoStop}%
\bibitem [{\citenamefont {Singh}\ \emph {et~al.}(2020)\citenamefont {Singh},
  \citenamefont {Herath}, \citenamefont {Wah}, \citenamefont {Liao},
  \citenamefont {Romero},\ and\ \citenamefont {Park}}]{SINGH2020107778}%
  \BibitemOpen
  \bibfield  {author} {\bibinfo {author} {\bibfnamefont {V.}~\bibnamefont
  {Singh}}, \bibinfo {author} {\bibfnamefont {U.}~\bibnamefont {Herath}},
  \bibinfo {author} {\bibfnamefont {B.}~\bibnamefont {Wah}}, \bibinfo {author}
  {\bibfnamefont {X.}~\bibnamefont {Liao}}, \bibinfo {author} {\bibfnamefont
  {A.~H.}\ \bibnamefont {Romero}},\ and\ \bibinfo {author} {\bibfnamefont
  {H.}~\bibnamefont {Park}},\ }\href
  {https://doi.org/https://doi.org/10.1016/j.cpc.2020.107778} {\bibfield
  {journal} {\bibinfo  {journal} {Computer Physics Communications}\ ,\ \bibinfo
  {pages} {107778}} (\bibinfo {year} {2020})}\BibitemShut {NoStop}%
\bibitem [{\citenamefont {Marzari}\ and\ \citenamefont
  {Vanderbilt}(1997)}]{MLWF1}%
  \BibitemOpen
  \bibfield  {author} {\bibinfo {author} {\bibfnamefont {N.}~\bibnamefont
  {Marzari}}\ and\ \bibinfo {author} {\bibfnamefont {D.}~\bibnamefont
  {Vanderbilt}},\ }\href {https://doi.org/10.1103/PhysRevB.56.12847} {\bibfield
   {journal} {\bibinfo  {journal} {Phys. Rev. B}\ }\textbf {\bibinfo {volume}
  {56}},\ \bibinfo {pages} {12847} (\bibinfo {year} {1997})}\BibitemShut
  {NoStop}%
\bibitem [{\citenamefont {Souza}\ \emph {et~al.}(2001)\citenamefont {Souza},
  \citenamefont {Marzari},\ and\ \citenamefont {Vanderbilt}}]{MLWF2}%
  \BibitemOpen
  \bibfield  {author} {\bibinfo {author} {\bibfnamefont {I.}~\bibnamefont
  {Souza}}, \bibinfo {author} {\bibfnamefont {N.}~\bibnamefont {Marzari}},\
  and\ \bibinfo {author} {\bibfnamefont {D.}~\bibnamefont {Vanderbilt}},\
  }\href {https://doi.org/10.1103/PhysRevB.65.035109} {\bibfield  {journal}
  {\bibinfo  {journal} {Phys. Rev. B}\ }\textbf {\bibinfo {volume} {65}},\
  \bibinfo {pages} {035109} (\bibinfo {year} {2001})}\BibitemShut {NoStop}%
\bibitem [{\citenamefont {Ruppen}\ \emph {et~al.}(2015)\citenamefont {Ruppen},
  \citenamefont {Teyssier}, \citenamefont {Peil}, \citenamefont {Catalano},
  \citenamefont {Gibert}, \citenamefont {Mravlje}, \citenamefont {Triscone},
  \citenamefont {Georges},\ and\ \citenamefont {van~der
  Marel}}]{PhysRevB.92.155145}%
  \BibitemOpen
  \bibfield  {author} {\bibinfo {author} {\bibfnamefont {J.}~\bibnamefont
  {Ruppen}}, \bibinfo {author} {\bibfnamefont {J.}~\bibnamefont {Teyssier}},
  \bibinfo {author} {\bibfnamefont {O.~E.}\ \bibnamefont {Peil}}, \bibinfo
  {author} {\bibfnamefont {S.}~\bibnamefont {Catalano}}, \bibinfo {author}
  {\bibfnamefont {M.}~\bibnamefont {Gibert}}, \bibinfo {author} {\bibfnamefont
  {J.}~\bibnamefont {Mravlje}}, \bibinfo {author} {\bibfnamefont {J.-M.}\
  \bibnamefont {Triscone}}, \bibinfo {author} {\bibfnamefont {A.}~\bibnamefont
  {Georges}},\ and\ \bibinfo {author} {\bibfnamefont {D.}~\bibnamefont {van~der
  Marel}},\ }\href {https://doi.org/10.1103/PhysRevB.92.155145} {\bibfield
  {journal} {\bibinfo  {journal} {Phys. Rev. B}\ }\textbf {\bibinfo {volume}
  {92}},\ \bibinfo {pages} {155145} (\bibinfo {year} {2015})}\BibitemShut
  {NoStop}%
\bibitem [{\citenamefont {Hampel}\ \emph {et~al.}(2019)\citenamefont {Hampel},
  \citenamefont {Liu}, \citenamefont {Franchini},\ and\ \citenamefont
  {Ederer}}]{Hampel19}%
  \BibitemOpen
  \bibfield  {author} {\bibinfo {author} {\bibfnamefont {A.}~\bibnamefont
  {Hampel}}, \bibinfo {author} {\bibfnamefont {P.}~\bibnamefont {Liu}},
  \bibinfo {author} {\bibfnamefont {C.}~\bibnamefont {Franchini}},\ and\
  \bibinfo {author} {\bibfnamefont {C.}~\bibnamefont {Ederer}},\ }\href@noop {}
  {\bibfield  {journal} {\bibinfo  {journal} {npj Quantum Materials}\ }\textbf
  {\bibinfo {volume} {4}},\ \bibinfo {pages} {5} (\bibinfo {year}
  {2019})}\BibitemShut {NoStop}%
\bibitem [{\citenamefont {Park}\ \emph
  {et~al.}(2014{\natexlab{a}})\citenamefont {Park}, \citenamefont {Millis},\
  and\ \citenamefont {Marianetti}}]{phase_LNO3}%
  \BibitemOpen
  \bibfield  {author} {\bibinfo {author} {\bibfnamefont {H.}~\bibnamefont
  {Park}}, \bibinfo {author} {\bibfnamefont {A.~J.}\ \bibnamefont {Millis}},\
  and\ \bibinfo {author} {\bibfnamefont {C.~A.}\ \bibnamefont {Marianetti}},\
  }\href {https://doi.org/10.1103/PhysRevB.89.245133} {\bibfield  {journal}
  {\bibinfo  {journal} {Phys. Rev. B}\ }\textbf {\bibinfo {volume} {89}},\
  \bibinfo {pages} {245133} (\bibinfo {year} {2014}{\natexlab{a}})}\BibitemShut
  {NoStop}%
\bibitem [{\citenamefont {Nowadnick}\ \emph {et~al.}(2015)\citenamefont
  {Nowadnick}, \citenamefont {Ruf}, \citenamefont {Park}, \citenamefont {King},
  \citenamefont {Schlom}, \citenamefont {Shen},\ and\ \citenamefont
  {Millis}}]{ARPES_DMFT}%
  \BibitemOpen
  \bibfield  {author} {\bibinfo {author} {\bibfnamefont {E.~A.}\ \bibnamefont
  {Nowadnick}}, \bibinfo {author} {\bibfnamefont {J.~P.}\ \bibnamefont {Ruf}},
  \bibinfo {author} {\bibfnamefont {H.}~\bibnamefont {Park}}, \bibinfo {author}
  {\bibfnamefont {P.~D.~C.}\ \bibnamefont {King}}, \bibinfo {author}
  {\bibfnamefont {D.~G.}\ \bibnamefont {Schlom}}, \bibinfo {author}
  {\bibfnamefont {K.~M.}\ \bibnamefont {Shen}},\ and\ \bibinfo {author}
  {\bibfnamefont {A.~J.}\ \bibnamefont {Millis}},\ }\href
  {https://doi.org/10.1103/PhysRevB.92.245109} {\bibfield  {journal} {\bibinfo
  {journal} {Phys. Rev. B}\ }\textbf {\bibinfo {volume} {92}},\ \bibinfo
  {pages} {245109} (\bibinfo {year} {2015})}\BibitemShut {NoStop}%
\bibitem [{\citenamefont {Gou}\ \emph {et~al.}(2011)\citenamefont {Gou},
  \citenamefont {Grinberg}, \citenamefont {Rappe},\ and\ \citenamefont
  {Rondinelli}}]{PhysRevB.84.144101}%
  \BibitemOpen
  \bibfield  {author} {\bibinfo {author} {\bibfnamefont {G.}~\bibnamefont
  {Gou}}, \bibinfo {author} {\bibfnamefont {I.}~\bibnamefont {Grinberg}},
  \bibinfo {author} {\bibfnamefont {A.~M.}\ \bibnamefont {Rappe}},\ and\
  \bibinfo {author} {\bibfnamefont {J.~M.}\ \bibnamefont {Rondinelli}},\ }\href
  {https://doi.org/10.1103/PhysRevB.84.144101} {\bibfield  {journal} {\bibinfo
  {journal} {Phys. Rev. B}\ }\textbf {\bibinfo {volume} {84}},\ \bibinfo
  {pages} {144101} (\bibinfo {year} {2011})}\BibitemShut {NoStop}%
\bibitem [{\citenamefont {Park}\ \emph
  {et~al.}(2014{\natexlab{b}})\citenamefont {Park}, \citenamefont {Millis},\
  and\ \citenamefont {Marianetti}}]{Park_dc_2014}%
  \BibitemOpen
  \bibfield  {author} {\bibinfo {author} {\bibfnamefont {H.}~\bibnamefont
  {Park}}, \bibinfo {author} {\bibfnamefont {A.~J.}\ \bibnamefont {Millis}},\
  and\ \bibinfo {author} {\bibfnamefont {C.~A.}\ \bibnamefont {Marianetti}},\
  }\href {https://doi.org/10.1103/PhysRevB.90.235103} {\bibfield  {journal}
  {\bibinfo  {journal} {Phys. Rev. B}\ }\textbf {\bibinfo {volume} {90}},\
  \bibinfo {pages} {235103} (\bibinfo {year} {2014}{\natexlab{b}})}\BibitemShut
  {NoStop}%
\bibitem [{\citenamefont {Gull}\ \emph {et~al.}(2011)\citenamefont {Gull},
  \citenamefont {Millis}, \citenamefont {Lichtenstein}, \citenamefont
  {Rubtsov}, \citenamefont {Troyer},\ and\ \citenamefont {Werner}}]{CTQMC}%
  \BibitemOpen
  \bibfield  {author} {\bibinfo {author} {\bibfnamefont {E.}~\bibnamefont
  {Gull}}, \bibinfo {author} {\bibfnamefont {A.~J.}\ \bibnamefont {Millis}},
  \bibinfo {author} {\bibfnamefont {A.~I.}\ \bibnamefont {Lichtenstein}},
  \bibinfo {author} {\bibfnamefont {A.~N.}\ \bibnamefont {Rubtsov}}, \bibinfo
  {author} {\bibfnamefont {M.}~\bibnamefont {Troyer}},\ and\ \bibinfo {author}
  {\bibfnamefont {P.}~\bibnamefont {Werner}},\ }\href
  {https://doi.org/10.1103/RevModPhys.83.349} {\bibfield  {journal} {\bibinfo
  {journal} {Rev. Mod. Phys.}\ }\textbf {\bibinfo {volume} {83}},\ \bibinfo
  {pages} {349} (\bibinfo {year} {2011})}\BibitemShut {NoStop}%
\bibitem [{\citenamefont {Haule}(2007)}]{PhysRevB.75.155113}%
  \BibitemOpen
  \bibfield  {author} {\bibinfo {author} {\bibfnamefont {K.}~\bibnamefont
  {Haule}},\ }\href {https://doi.org/10.1103/PhysRevB.75.155113} {\bibfield
  {journal} {\bibinfo  {journal} {Phys. Rev. B}\ }\textbf {\bibinfo {volume}
  {75}},\ \bibinfo {pages} {155113} (\bibinfo {year} {2007})}\BibitemShut
  {NoStop}%
\bibitem [{\citenamefont {Wold}\ \emph {et~al.}(1957)\citenamefont {Wold},
  \citenamefont {Post},\ and\ \citenamefont {Banks}}]{LNO3_rhom}%
  \BibitemOpen
  \bibfield  {author} {\bibinfo {author} {\bibfnamefont {A.}~\bibnamefont
  {Wold}}, \bibinfo {author} {\bibfnamefont {B.}~\bibnamefont {Post}},\ and\
  \bibinfo {author} {\bibfnamefont {E.}~\bibnamefont {Banks}},\ }\href
  {https://doi.org/10.1021/ja01575a022} {\bibfield  {journal} {\bibinfo
  {journal} {Journal of the American Chemical Society}\ }\textbf {\bibinfo
  {volume} {79}},\ \bibinfo {pages} {4911} (\bibinfo {year} {1957})},\ \Eprint
  {https://arxiv.org/abs/https://doi.org/10.1021/ja01575a022}
  {https://doi.org/10.1021/ja01575a022} \BibitemShut {NoStop}%
\bibitem [{\citenamefont {Alonso}\ \emph {et~al.}(1997)\citenamefont {Alonso},
  \citenamefont {Mart{\'{\i}}nez-Lope}, \citenamefont
  {Garc{\'{\i}}a-Mu{\~{n}}oz},\ and\ \citenamefont
  {Fern{\'{a}}ndez-D{\'{\i}}az}}]{Alonso_1997}%
  \BibitemOpen
  \bibfield  {author} {\bibinfo {author} {\bibfnamefont {J.~A.}\ \bibnamefont
  {Alonso}}, \bibinfo {author} {\bibfnamefont {M.~J.}\ \bibnamefont
  {Mart{\'{\i}}nez-Lope}}, \bibinfo {author} {\bibfnamefont {J.~L.}\
  \bibnamefont {Garc{\'{\i}}a-Mu{\~{n}}oz}},\ and\ \bibinfo {author}
  {\bibfnamefont {M.~T.}\ \bibnamefont {Fern{\'{a}}ndez-D{\'{\i}}az}},\ }\href
  {https://doi.org/10.1088/0953-8984/9/30/010} {\bibfield  {journal} {\bibinfo
  {journal} {Journal of Physics: Condensed Matter}\ }\textbf {\bibinfo {volume}
  {9}},\ \bibinfo {pages} {6417} (\bibinfo {year} {1997})}\BibitemShut
  {NoStop}%
\bibitem [{\citenamefont {Moriga}\ \emph {et~al.}(1994)\citenamefont {Moriga},
  \citenamefont {Usaka}, \citenamefont {Imamura}, \citenamefont {Nakabayashi},
  \citenamefont {Matsubara}, \citenamefont {Kinouchi}, \citenamefont
  {Shinichi},\ and\ \citenamefont {Kanamaru}}]{Moriga_structure}%
  \BibitemOpen
  \bibfield  {author} {\bibinfo {author} {\bibfnamefont {T.}~\bibnamefont
  {Moriga}}, \bibinfo {author} {\bibfnamefont {O.}~\bibnamefont {Usaka}},
  \bibinfo {author} {\bibfnamefont {T.}~\bibnamefont {Imamura}}, \bibinfo
  {author} {\bibfnamefont {I.}~\bibnamefont {Nakabayashi}}, \bibinfo {author}
  {\bibfnamefont {I.}~\bibnamefont {Matsubara}}, \bibinfo {author}
  {\bibfnamefont {T.}~\bibnamefont {Kinouchi}}, \bibinfo {author}
  {\bibfnamefont {K.}~\bibnamefont {Shinichi}},\ and\ \bibinfo {author}
  {\bibfnamefont {F.}~\bibnamefont {Kanamaru}},\ }\href
  {https://doi.org/10.1246/bcsj.67.687} {\bibfield  {journal} {\bibinfo
  {journal} {Bulletin of the Chemical Society of Japan}\ }\textbf {\bibinfo
  {volume} {67}},\ \bibinfo {pages} {687} (\bibinfo {year} {1994})}\BibitemShut
  {NoStop}%
\bibitem [{\citenamefont {Garc\'{\i}a-Mu\~noz}\ \emph
  {et~al.}(1992)\citenamefont {Garc\'{\i}a-Mu\~noz}, \citenamefont
  {Rodr\'{\i}guez-Carvajal}, \citenamefont {Lacorre},\ and\ \citenamefont
  {Torrance}}]{LNO_LattParaExp}%
  \BibitemOpen
  \bibfield  {author} {\bibinfo {author} {\bibfnamefont {J.~L.}\ \bibnamefont
  {Garc\'{\i}a-Mu\~noz}}, \bibinfo {author} {\bibfnamefont {J.}~\bibnamefont
  {Rodr\'{\i}guez-Carvajal}}, \bibinfo {author} {\bibfnamefont
  {P.}~\bibnamefont {Lacorre}},\ and\ \bibinfo {author} {\bibfnamefont {J.~B.}\
  \bibnamefont {Torrance}},\ }\href {https://doi.org/10.1103/PhysRevB.46.4414}
  {\bibfield  {journal} {\bibinfo  {journal} {Phys. Rev. B}\ }\textbf {\bibinfo
  {volume} {46}},\ \bibinfo {pages} {4414} (\bibinfo {year}
  {1992})}\BibitemShut {NoStop}%
\bibitem [{\citenamefont {Geisler}\ and\ \citenamefont
  {Pentcheva}(2020)}]{Geisler}%
  \BibitemOpen
  \bibfield  {author} {\bibinfo {author} {\bibfnamefont {B.}~\bibnamefont
  {Geisler}}\ and\ \bibinfo {author} {\bibfnamefont {R.}~\bibnamefont
  {Pentcheva}},\ }\href {https://doi.org/10.1103/PhysRevB.101.165108}
  {\bibfield  {journal} {\bibinfo  {journal} {Phys. Rev. B}\ }\textbf {\bibinfo
  {volume} {101}},\ \bibinfo {pages} {165108} (\bibinfo {year}
  {2020})}\BibitemShut {NoStop}%
\bibitem [{\citenamefont {Reuter}\ and\ \citenamefont
  {Scheffler}(2001)}]{For1}%
  \BibitemOpen
  \bibfield  {author} {\bibinfo {author} {\bibfnamefont {K.}~\bibnamefont
  {Reuter}}\ and\ \bibinfo {author} {\bibfnamefont {M.}~\bibnamefont
  {Scheffler}},\ }\href {https://doi.org/10.1103/PhysRevB.65.035406} {\bibfield
   {journal} {\bibinfo  {journal} {Phys. Rev. B}\ }\textbf {\bibinfo {volume}
  {65}},\ \bibinfo {pages} {035406} (\bibinfo {year} {2001})}\BibitemShut
  {NoStop}%
\bibitem [{\citenamefont {Pentcheva}\ \emph {et~al.}(2005)\citenamefont
  {Pentcheva}, \citenamefont {Wendler}, \citenamefont {Meyerheim},
  \citenamefont {Moritz}, \citenamefont {Jedrecy},\ and\ \citenamefont
  {Scheffler}}]{For2}%
  \BibitemOpen
  \bibfield  {author} {\bibinfo {author} {\bibfnamefont {R.}~\bibnamefont
  {Pentcheva}}, \bibinfo {author} {\bibfnamefont {F.}~\bibnamefont {Wendler}},
  \bibinfo {author} {\bibfnamefont {H.~L.}\ \bibnamefont {Meyerheim}}, \bibinfo
  {author} {\bibfnamefont {W.}~\bibnamefont {Moritz}}, \bibinfo {author}
  {\bibfnamefont {N.}~\bibnamefont {Jedrecy}},\ and\ \bibinfo {author}
  {\bibfnamefont {M.}~\bibnamefont {Scheffler}},\ }\href
  {https://doi.org/10.1103/PhysRevLett.94.126101} {\bibfield  {journal}
  {\bibinfo  {journal} {Phys. Rev. Lett.}\ }\textbf {\bibinfo {volume} {94}},\
  \bibinfo {pages} {126101} (\bibinfo {year} {2005})}\BibitemShut {NoStop}%
\bibitem [{\citenamefont {Mulakaluri}\ \emph {et~al.}(2009)\citenamefont
  {Mulakaluri}, \citenamefont {Pentcheva}, \citenamefont {Wieland},
  \citenamefont {Moritz},\ and\ \citenamefont {Scheffler}}]{For3}%
  \BibitemOpen
  \bibfield  {author} {\bibinfo {author} {\bibfnamefont {N.}~\bibnamefont
  {Mulakaluri}}, \bibinfo {author} {\bibfnamefont {R.}~\bibnamefont
  {Pentcheva}}, \bibinfo {author} {\bibfnamefont {M.}~\bibnamefont {Wieland}},
  \bibinfo {author} {\bibfnamefont {W.}~\bibnamefont {Moritz}},\ and\ \bibinfo
  {author} {\bibfnamefont {M.}~\bibnamefont {Scheffler}},\ }\href
  {https://doi.org/10.1103/PhysRevLett.103.176102} {\bibfield  {journal}
  {\bibinfo  {journal} {Phys. Rev. Lett.}\ }\textbf {\bibinfo {volume} {103}},\
  \bibinfo {pages} {176102} (\bibinfo {year} {2009})}\BibitemShut {NoStop}%
\bibitem [{\citenamefont {Hepting}\ \emph {et~al.}(2020)\citenamefont
  {Hepting}, \citenamefont {Li}, \citenamefont {Jia}, \citenamefont {Lu},
  \citenamefont {Paris}, \citenamefont {Tseng}, \citenamefont {Feng},
  \citenamefont {Osada}, \citenamefont {Been}, \citenamefont {Hikita},
  \citenamefont {Chuang}, \citenamefont {Hussain}, \citenamefont {Zhou},
  \citenamefont {Nag}, \citenamefont {Garcia-Fernandez}, \citenamefont {Rossi},
  \citenamefont {Huang}, \citenamefont {Huang}, \citenamefont {Shen},
  \citenamefont {Schmitt}, \citenamefont {Hwang}, \citenamefont {Moritz},
  \citenamefont {Zaanen}, \citenamefont {Devereaux},\ and\ \citenamefont
  {Lee}}]{LNO_NNO_sc}%
  \BibitemOpen
  \bibfield  {author} {\bibinfo {author} {\bibfnamefont {M.}~\bibnamefont
  {Hepting}}, \bibinfo {author} {\bibfnamefont {D.}~\bibnamefont {Li}},
  \bibinfo {author} {\bibfnamefont {C.~J.}\ \bibnamefont {Jia}}, \bibinfo
  {author} {\bibfnamefont {H.}~\bibnamefont {Lu}}, \bibinfo {author}
  {\bibfnamefont {E.}~\bibnamefont {Paris}}, \bibinfo {author} {\bibfnamefont
  {Y.}~\bibnamefont {Tseng}}, \bibinfo {author} {\bibfnamefont
  {X.}~\bibnamefont {Feng}}, \bibinfo {author} {\bibfnamefont {M.}~\bibnamefont
  {Osada}}, \bibinfo {author} {\bibfnamefont {E.}~\bibnamefont {Been}},
  \bibinfo {author} {\bibfnamefont {Y.}~\bibnamefont {Hikita}}, \bibinfo
  {author} {\bibfnamefont {Y.-D.}\ \bibnamefont {Chuang}}, \bibinfo {author}
  {\bibfnamefont {Z.}~\bibnamefont {Hussain}}, \bibinfo {author} {\bibfnamefont
  {K.~J.}\ \bibnamefont {Zhou}}, \bibinfo {author} {\bibfnamefont
  {A.}~\bibnamefont {Nag}}, \bibinfo {author} {\bibfnamefont {M.}~\bibnamefont
  {Garcia-Fernandez}}, \bibinfo {author} {\bibfnamefont {M.}~\bibnamefont
  {Rossi}}, \bibinfo {author} {\bibfnamefont {H.~Y.}\ \bibnamefont {Huang}},
  \bibinfo {author} {\bibfnamefont {D.~J.}\ \bibnamefont {Huang}}, \bibinfo
  {author} {\bibfnamefont {Z.~X.}\ \bibnamefont {Shen}}, \bibinfo {author}
  {\bibfnamefont {T.}~\bibnamefont {Schmitt}}, \bibinfo {author} {\bibfnamefont
  {H.~Y.}\ \bibnamefont {Hwang}}, \bibinfo {author} {\bibfnamefont
  {B.}~\bibnamefont {Moritz}}, \bibinfo {author} {\bibfnamefont
  {J.}~\bibnamefont {Zaanen}}, \bibinfo {author} {\bibfnamefont {T.~P.}\
  \bibnamefont {Devereaux}},\ and\ \bibinfo {author} {\bibfnamefont {W.~S.}\
  \bibnamefont {Lee}},\ }\href
  {https://doi.org/https://doi.org/10.1038/s41563-019-0585-z} {\bibfield
  {journal} {\bibinfo  {journal} {Nat. Mater.}\ }\textbf {\bibinfo {volume}
  {19}},\ \bibinfo {pages} {381} (\bibinfo {year} {2020})}\BibitemShut
  {NoStop}%
\bibitem [{\citenamefont {Kurmaev}\ \emph {et~al.}(2008)\citenamefont
  {Kurmaev}, \citenamefont {Wilks}, \citenamefont {Moewes}, \citenamefont
  {Finkelstein}, \citenamefont {Shamin},\ and\ \citenamefont
  {Kune\ifmmode~\check{s}\else \v{s}\fi{}}}]{PhysRevB.77.165127}%
  \BibitemOpen
  \bibfield  {author} {\bibinfo {author} {\bibfnamefont {E.~Z.}\ \bibnamefont
  {Kurmaev}}, \bibinfo {author} {\bibfnamefont {R.~G.}\ \bibnamefont {Wilks}},
  \bibinfo {author} {\bibfnamefont {A.}~\bibnamefont {Moewes}}, \bibinfo
  {author} {\bibfnamefont {L.~D.}\ \bibnamefont {Finkelstein}}, \bibinfo
  {author} {\bibfnamefont {S.~N.}\ \bibnamefont {Shamin}},\ and\ \bibinfo
  {author} {\bibfnamefont {J.}~\bibnamefont {Kune\ifmmode~\check{s}\else
  \v{s}\fi{}}},\ }\href {https://doi.org/10.1103/PhysRevB.77.165127} {\bibfield
   {journal} {\bibinfo  {journal} {Phys. Rev. B}\ }\textbf {\bibinfo {volume}
  {77}},\ \bibinfo {pages} {165127} (\bibinfo {year} {2008})}\BibitemShut
  {NoStop}%
\bibitem [{\citenamefont {Timrov}\ \emph {et~al.}(2020)\citenamefont {Timrov},
  \citenamefont {Agrawal}, \citenamefont {Zhang}, \citenamefont {Erat},
  \citenamefont {Liu}, \citenamefont {Braun}, \citenamefont {Cococcioni},
  \citenamefont {Calandra}, \citenamefont {Marzari},\ and\ \citenamefont
  {Passerone}}]{PhysRevResearch.2.033265}%
  \BibitemOpen
  \bibfield  {author} {\bibinfo {author} {\bibfnamefont {I.}~\bibnamefont
  {Timrov}}, \bibinfo {author} {\bibfnamefont {P.}~\bibnamefont {Agrawal}},
  \bibinfo {author} {\bibfnamefont {X.}~\bibnamefont {Zhang}}, \bibinfo
  {author} {\bibfnamefont {S.}~\bibnamefont {Erat}}, \bibinfo {author}
  {\bibfnamefont {R.}~\bibnamefont {Liu}}, \bibinfo {author} {\bibfnamefont
  {A.}~\bibnamefont {Braun}}, \bibinfo {author} {\bibfnamefont
  {M.}~\bibnamefont {Cococcioni}}, \bibinfo {author} {\bibfnamefont
  {M.}~\bibnamefont {Calandra}}, \bibinfo {author} {\bibfnamefont
  {N.}~\bibnamefont {Marzari}},\ and\ \bibinfo {author} {\bibfnamefont
  {D.}~\bibnamefont {Passerone}},\ }\href
  {https://doi.org/10.1103/PhysRevResearch.2.033265} {\bibfield  {journal}
  {\bibinfo  {journal} {Phys. Rev. Research}\ }\textbf {\bibinfo {volume}
  {2}},\ \bibinfo {pages} {033265} (\bibinfo {year} {2020})}\BibitemShut
  {NoStop}%
\bibitem [{\citenamefont {Park}\ \emph {et~al.}(2020)\citenamefont {Park},
  \citenamefont {Nanguneri},\ and\ \citenamefont {Ngo}}]{LaCoO3_SSMT}%
  \BibitemOpen
  \bibfield  {author} {\bibinfo {author} {\bibfnamefont {H.}~\bibnamefont
  {Park}}, \bibinfo {author} {\bibfnamefont {R.}~\bibnamefont {Nanguneri}},\
  and\ \bibinfo {author} {\bibfnamefont {A.~T.}\ \bibnamefont {Ngo}},\ }\href
  {https://doi.org/10.1103/PhysRevB.101.195125} {\bibfield  {journal} {\bibinfo
   {journal} {Phys. Rev. B}\ }\textbf {\bibinfo {volume} {101}},\ \bibinfo
  {pages} {195125} (\bibinfo {year} {2020})}\BibitemShut {NoStop}%
\bibitem [{\citenamefont {Greenberg}\ \emph {et~al.}(2018)\citenamefont
  {Greenberg}, \citenamefont {Leonov}, \citenamefont {Layek}, \citenamefont
  {Konopkova}, \citenamefont {Pasternak}, \citenamefont {Dubrovinsky},
  \citenamefont {Jeanloz}, \citenamefont {Abrikosov},\ and\ \citenamefont
  {Rozenberg}}]{PhysRevX.8.031059}%
  \BibitemOpen
  \bibfield  {author} {\bibinfo {author} {\bibfnamefont {E.}~\bibnamefont
  {Greenberg}}, \bibinfo {author} {\bibfnamefont {I.}~\bibnamefont {Leonov}},
  \bibinfo {author} {\bibfnamefont {S.}~\bibnamefont {Layek}}, \bibinfo
  {author} {\bibfnamefont {Z.}~\bibnamefont {Konopkova}}, \bibinfo {author}
  {\bibfnamefont {M.~P.}\ \bibnamefont {Pasternak}}, \bibinfo {author}
  {\bibfnamefont {L.}~\bibnamefont {Dubrovinsky}}, \bibinfo {author}
  {\bibfnamefont {R.}~\bibnamefont {Jeanloz}}, \bibinfo {author} {\bibfnamefont
  {I.~A.}\ \bibnamefont {Abrikosov}},\ and\ \bibinfo {author} {\bibfnamefont
  {G.~K.}\ \bibnamefont {Rozenberg}},\ }\href
  {https://doi.org/10.1103/PhysRevX.8.031059} {\bibfield  {journal} {\bibinfo
  {journal} {Phys. Rev. X}\ }\textbf {\bibinfo {volume} {8}},\ \bibinfo {pages}
  {031059} (\bibinfo {year} {2018})}\BibitemShut {NoStop}%
\bibitem [{\citenamefont {Leonov}\ \emph {et~al.}(2019)\citenamefont {Leonov},
  \citenamefont {Rozenberg},\ and\ \citenamefont {Abrikosov}}]{Fe2O3_npj}%
  \BibitemOpen
  \bibfield  {author} {\bibinfo {author} {\bibfnamefont {I.}~\bibnamefont
  {Leonov}}, \bibinfo {author} {\bibfnamefont {G.~K.}\ \bibnamefont
  {Rozenberg}},\ and\ \bibinfo {author} {\bibfnamefont {I.~A.}\ \bibnamefont
  {Abrikosov}},\ }\bibfield  {journal} {\bibinfo  {journal} {npj Computational
  Materials}\ }\textbf {\bibinfo {volume} {5}},\ \href
  {https://doi.org/10.1038/s41524-019-0225-9} {10.1038/s41524-019-0225-9}
  (\bibinfo {year} {2019})\BibitemShut {NoStop}%
\bibitem [{\citenamefont {Valli}\ \emph {et~al.}(2015)\citenamefont {Valli},
  \citenamefont {Das}, \citenamefont {Sangiovanni}, \citenamefont
  {Saha-Dasgupta},\ and\ \citenamefont {Held}}]{PhysRevB.92.115143}%
  \BibitemOpen
  \bibfield  {author} {\bibinfo {author} {\bibfnamefont {A.}~\bibnamefont
  {Valli}}, \bibinfo {author} {\bibfnamefont {H.}~\bibnamefont {Das}}, \bibinfo
  {author} {\bibfnamefont {G.}~\bibnamefont {Sangiovanni}}, \bibinfo {author}
  {\bibfnamefont {T.}~\bibnamefont {Saha-Dasgupta}},\ and\ \bibinfo {author}
  {\bibfnamefont {K.}~\bibnamefont {Held}},\ }\href
  {https://doi.org/10.1103/PhysRevB.92.115143} {\bibfield  {journal} {\bibinfo
  {journal} {Phys. Rev. B}\ }\textbf {\bibinfo {volume} {92}},\ \bibinfo
  {pages} {115143} (\bibinfo {year} {2015})}\BibitemShut {NoStop}%
\end{thebibliography}%

\clearpage
\setcounter{page}{0}
\thispagestyle{empty}
\onecolumngrid

\section*{Supplementary materials}
\subsection{DFT and DFT+U density of states} 
\label{sec:DFTDOS}
\begin{figure}
    \begin{subfigure}[b]{0.3\linewidth}
       \centering
       \includegraphics[width=0.8\linewidth]{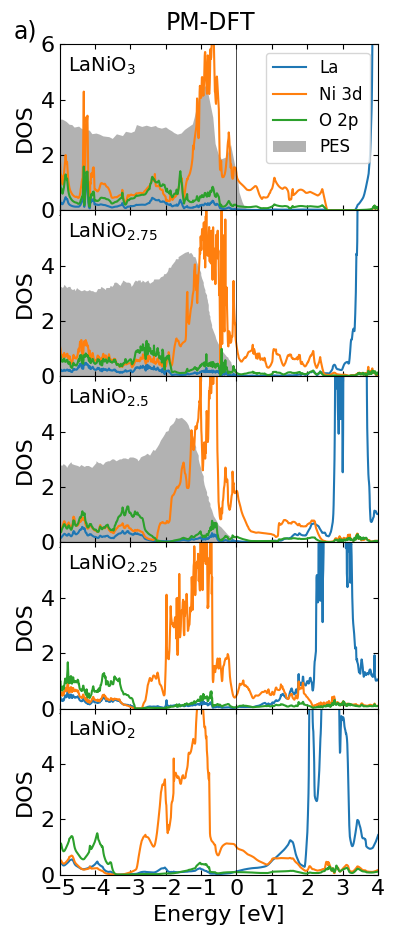}
    \end{subfigure}
    \begin{subfigure}[b]{0.3\linewidth}
       \centering
       \includegraphics[width=0.8\linewidth]{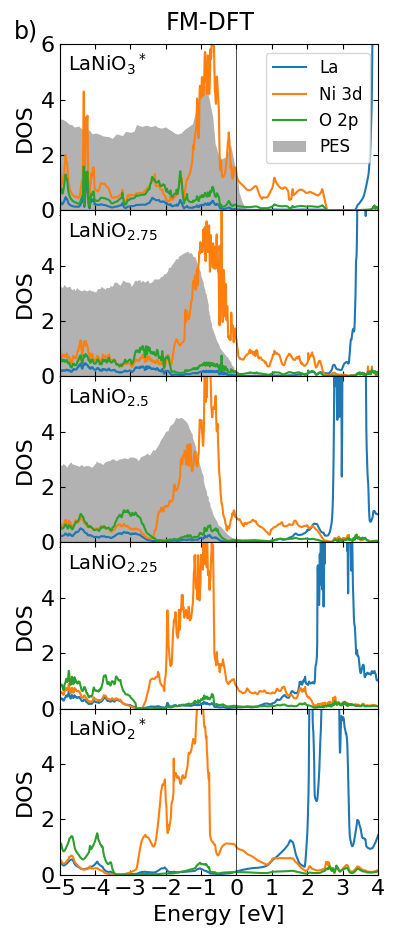}
    \end{subfigure}
    \begin{subfigure}[b]{0.3\linewidth}
       \centering
       \includegraphics[width=0.8\linewidth]{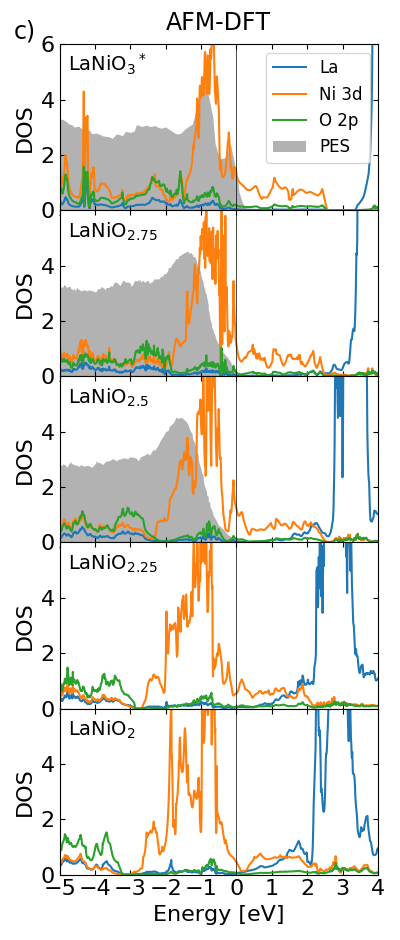}
    \end{subfigure}
    \begin{subfigure}[b]{0.3\linewidth}
       \centering
       \includegraphics[width=0.8\linewidth]{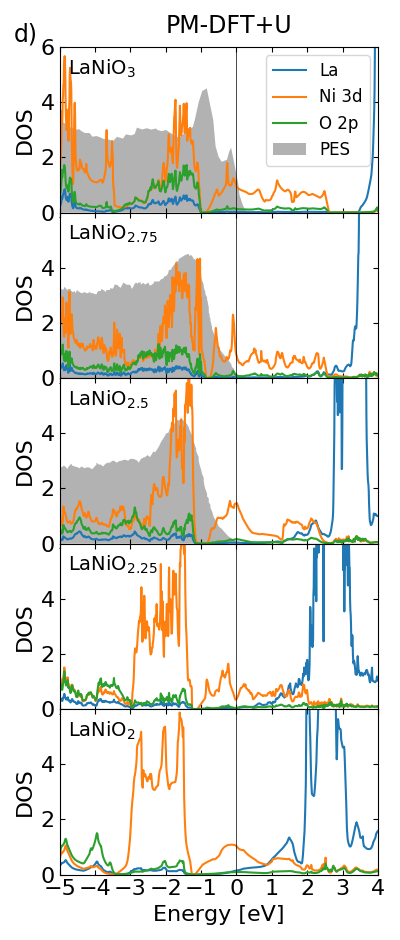}
    \end{subfigure}
    \begin{subfigure}[b]{0.3\linewidth}
       \centering
       \includegraphics[width=0.8\linewidth]{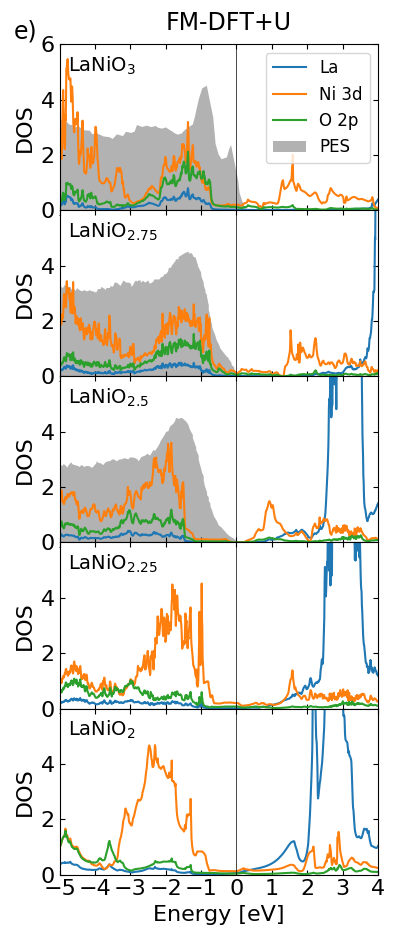}
    \end{subfigure}
    \begin{subfigure}[b]{0.3\linewidth}
       \centering
       \includegraphics[width=0.8\linewidth]{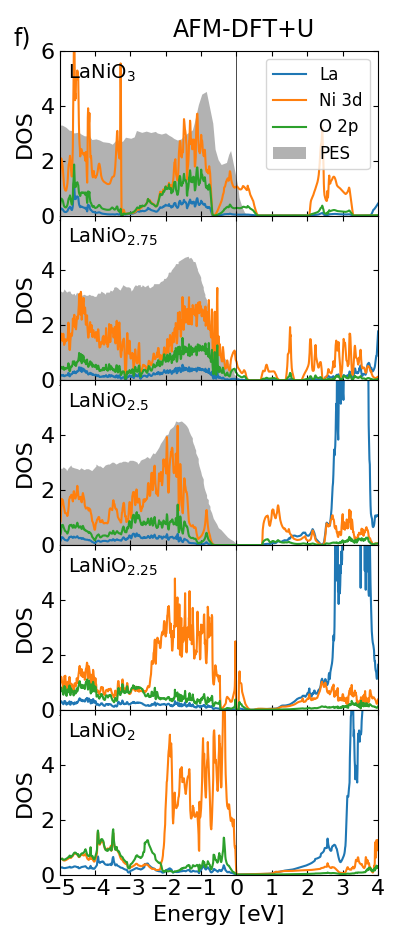}
    \end{subfigure}
\caption{Orbital-resolved density of states calculated using the DFT method for LaNiO$_{3}$, LaNiO$_{2.75}$, LaNiO$_{2.5}$, LaNiO$_{2.25}$ and LaNiO$_{2}$ with (a) non-magnetic, (b) ferromagnetic, and (c) antiferromagnetic orderings, as well as using the DFT+U method with (d) non-magnetic, (e) ferromagnetic, (f) antiferromagnetic orderings. The shaded region is taken from previous experimental work[22].
Sub-figures with * in label indicates the ground state converged to zero magnetic moment.}
\label{fig:dft_dos}
\end{figure}

In Fig.~\ref{fig:dft_dos}, we list the density of states (DOS) for LaNiO$_{3-x}$ with $x$=0, 0.25, 0.5, 0.75, and 1 computed using DFT and DFT+U with non-magnetic, ferromagnetic and G-type antiferromagnetic configurations. For LaNiO$_3$, LaNiO$_{2.75}$ and LaNiO$_{2.5}$ cases, we compare the orbital-resolved DOS with experimental photoemission spectroscopy (PES) spectra[22] as depicted in the shaded region. 
In all calculations, as the oxygen vacancy level $x$ increases, the La state moves down in energy close to the Fermi energy while the Ni state is slightly shifted below the Fermi energy. The Ni-O hybridization becomes also weaker as the vacancy level evolves.
The ground-states of all DFT calculations become metallic as DFT underestimates correlation effects.
In DFT+U, ferromagnetic and G-type antiferromagnetic ground states in LaNiO$_{2.5}$ become insulator consistently with the experimental transport property although the non-magnetic ground state is still metallic as the correlation effect is also underestimated.

\subsection{DFT+DMFT calculation details}
\label{sec:bandstructure}
\begin{figure}
    \begin{subfigure}[b]{1\linewidth}
       \centering
       \includegraphics[width=0.5\linewidth]{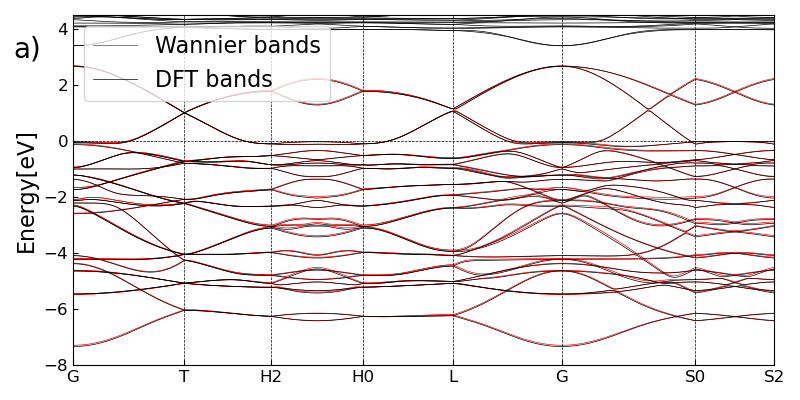}
    \end{subfigure}
    \begin{subfigure}[b]{1\linewidth}
       \centering
       \includegraphics[width=0.5\linewidth]{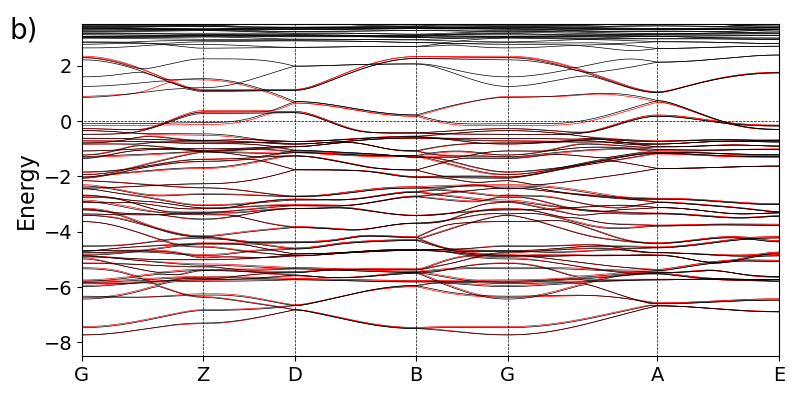}
    \end{subfigure}
    \begin{subfigure}[b]{1\linewidth}
        \centering
        \includegraphics[width=0.5\linewidth]{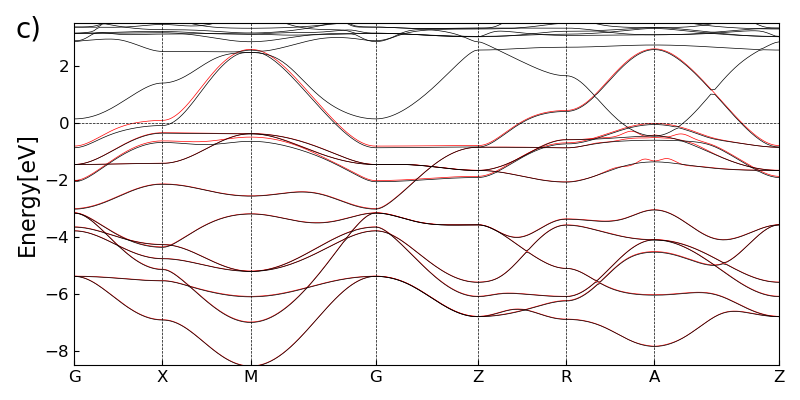}
    \end{subfigure}
\caption{Comparison of the Wannier band structures of Ni 3$d$ and O 2$p$ orbitals with the DFT band structures for (a) LaNiO$_{3}$, (b) LaNiO$_{2.5}$, and (c) LaNiO$_{2}$.}
\label{fig:bands_compare}
\end{figure}

The overall procedure of DFT+DMFT as implemented in the DMFTwDFT package[38] is as follows. 
First, we perform non-magnetic DFT calculations for each structure. 
Then, we adopt the Wannier90 package[39,40] to obtain maximally localized Wannier functions (MLWFs) as localized orbitals for DMFT.
To construct the Wannier orbitals, we take an energy window from -9 eV to +5 eV, with respect to Fermi energies, which basically contains all Ni 3$d$ and O 2$p$ orbitals. 
It is also important that the interpolated band structure obtained from MLWFs matches to the original DFT band structure. In Fig.$\:$\ref{fig:bands_compare}, we plot Wannier band structures for LaNiO$_{3}$, LaNiO$_{2.5}$ and LaNiO$_{2}$ and compare to DFT band structures. The Wannier bands match to the DFT bands almost perfectly capturing the original band structure in DFT.
We applied frozen window in MLWFs construction processes of LaNiO$_{2}$ and LaNiO$_{2.5}$. For LaNiO$_{2}$ we set frozen window to be -9 eV $\sim$ -1.56 eV with respect to the Fermi energy. For LaNiO$_{2.5}$ it is -7.9 eV $\sim$ 1.2 eV with respect to the Fermi energy. Without frozen windows setting, the Wannier bands have relatively large discrepancies with DFT bands for LaNiO$_{2}$ and LaNiO$_{2.5}$.

From the obtained Wannier subspace, we perform the DMFT self-consistent calculation using the continuous-time quantum Monte Carlo as an impurity solver.
We use  18$\times$18$\times$18 $k-$points which are denser than the DFT $k-$points while doing the DMFT calculations.
To minimize the off-diagonal component of hybridization functions in DMFT, we adopt the rotated axis for constructing MLWFs which is aligned to the bonding direction of the local Ni octahedron sites.
To avoid the double counting of the Coulomb interaction taken in DFT+DMFT, we also used the modified double counting correction energy (DC\_type = 1 in the DMFTwDFT package) with the parameter $\alpha$ = 0.2, which is given by
\begin{equation}
E^{DC}=\frac{(U-\alpha)}{2}{\cdot}N_d{\cdot}{(N_d-1)}-\frac{J}{4}{\cdot}N_d{\cdot}{(N_d-2)}
\end{equation}
where $U$ is the Hubbard interaction, $J$ is the Hund's coupling, and $N_d$ is the $d-$occupancy.

\subsection{Total energies of LaNiO$_{3-x}$}
\label{sec:total_energies}

In Table \ref{tbl:total_eng_all}, we list total energies of LaNiO$_{3}$, LaNiO$_{2.75}$, LaNiO$_{2.5}$, LaNiO$_{2.25}$, and LaNiO$_{2}$ obtained from DFT and DFT+U calculations by relaxing structures with different magnetism including PM, FM, and G-type AFM. The energy difference between FM and AFM for each structure was given and explained in the main text.
\begin{table*}[ht]
    \centering
    \caption{Total energies [eV] of LaNiO$_{3}$, LaNiO$_{2.75}$, LaNiO$_{2.5}$, LaNiO$_{2.25}$, and LaNiO$_{2}$} 
    \begin{ruledtabular}
    \begin{tabular}{p{0.1\linewidth}p{0.1\linewidth}p{0.1\linewidth}p{0.1\linewidth}p{0.03\linewidth}p{0.1\linewidth}p{0.1\linewidth}p{0.1\linewidth}}
       & & DFT & && & DFT+U & \\ 
       & PM & FM & AFM && PM & FM & AFM\\
    \hline
      LaNiO$_3$ &-36.041 & -36.041\footnotemark[1] & -36.041\footnotemark[1] && -33.852 &-34.175 &-33.834 \\ 
      LaNiO$_{2.75}$ &-34.437 &-34.442 &-34.438 &&  -32.160 &-32.650 &-32.610 \\ 
      LaNiO$_{2.5}$ &-32.678 &-32.723 &-32.713 &&  -30.460 &-31.098 &-31.139\\ 
      LaNiO$_{2.25}$ &-30.950 &-30.953 &-30.951 &&  -28.664 &-29.124 &-29.056\\ 
      LaNiO$_{2}$ &-29.233 & -29.233\footnotemark[1] &-29.213 &&  -26.842 & -27.396 &-27.447 \\ 
    \end{tabular}
    \footnotetext[1]{Relaxation converged to 0 magnetic momentum.}
    \end{ruledtabular}
\label{tbl:total_eng_all}
\end{table*}

\end{document}